\documentclass[12pt,preprint,showpacs,preprintnumbers]{revtex4}
\usepackage{amsfonts}
\usepackage{amsmath}
\usepackage{amssymb}
\usepackage{hyperref}
\usepackage{color}
\usepackage{graphicx}
\usepackage{amssymb}
\usepackage{amsmath}
\usepackage{graphicx}
\usepackage{dcolumn}
\usepackage{bm}
\usepackage{epsfig}
\usepackage[T1]{fontenc}
\usepackage{ae,aecompl}

\begin{document}

\title{Multiple-relaxation-time lattice Boltzmann kinetic model for combustion}
\author{ Aiguo Xu$^{1,2,3}$\footnote{Corresponding author. E-mail address: Xu\_Aiguo@iapcm.ac.cn},
Chuandong Lin$^{4}$,
Guangcai Zhang$^{1,3,5}$,
Yingjun Li$^{4}$\footnote{Corresponding author. E-mail address: lyj@aphy.iphy.ac.cn}
}
\affiliation{$^1$ National Key Laboratory of Computational Physics, Institute of Applied
Physics and Computational Mathematics, P. O. Box 8009-26, Beijing 100088,
P.R.China \\
$^2$ Center for Applied Physics and Technology, MOE Key Center for High
Energy Density Physics Simulations, College of Engineering, Peking
University, Beijing 100871, P.R.China \\
$^3$ State Key Laboratory of Explosion Science and Technology, Beijing
Institute of Technology, Beijing 100081, P.R.China \\
$^4$ State Key Laboratory for GeoMechanics and Deep Underground Engineering,
China University of Mining and Technology, Beijing 100083, P.R.China \\
$^5$ State Key Laboratory of Theoretical Physics, Institute of Theoretical
Physics, Chinese Academy of Sciences,Beijing 100190, P.R.China
}
\date{\today }

\begin{abstract}
To probe both the Hydrodynamic Non-Equilibrium (HNE) and Thermodynamic Non-Equilibrium (TNE) in the combustion process, a two-dimensional Multiple-Relaxation-Time (MRT) version of Lattice Boltzmann Kinetic Model(LBKM) for combustion phenomena is presented. The chemical energy released in the progress of combustion is dynamically coupled into the system by adding a chemical term to the LB kinetic equation.
Beside describing the evolutions of the conserved quantities, the density, momentum and energy, which are what the Navier-Stokes model describes, the MRT-LBKM presents also a coarse-grained description on the evolutions of some non-conserved quantities.
The current model works for both subsonic and supersonic flows with or without chemical reaction. In this model both the specific-heat ratio and the Prandtl number are flexible, the TNE effects are naturally presented in each simulation step.  The model is verified and validated via well-known benchmark tests.
As an initial application, various non-equilibrium behaviours, including the complex interplays between various HNEs, between various TNEs and between the HNE and TNE,
around the detonation wave in the unsteady and steady one-dimensional detonation processes are preliminarily  probed.
  It is found that the system viscosity (or heat conductivity) decreases the local TNE, but increase the global TNE around the detonation wave, that even locally, the system viscosity (or heat conductivity) results in two kinds of competing trends, to increase and to decrease the TNE effects.
The physical reason is that the viscosity (or heat conductivity) takes part in both the thermodynamic and hydrodynamic responses.
\end{abstract}

\maketitle

\section{Introduction}
Combustion has long been playing a dominant role in the transportation and power generation. More than $80\%$ of world energy is from various combustion processes. For a foreseeable future it will remain to be the major energy conversion process \cite{Ju-Review2014}. At the same time, the low energy conversion efficiency of existing combustion engines has been becoming the major source of air pollution and driving force for climate change \cite{Chu-Nature2012}. Roughly speaking, there are two kinds of fuels, the nuclear fuel and the organic fuel. The latter contains the organic materials such as hydrocarbon natural fuel and artificial fuel after processing. Various medical wastes \cite{Jangsawang2005} belong to the organic fuel. To achieve low emissions, fuel lean and high speed combustion and enable new engine technologies, in recent years, some new combustion concepts, such as pulsed and spinning detonation engines \cite{Schott1965,Bykovskii2006},
microscale combustion \cite{Ju2011,Fernandez-Pello2002} and nanopropellants \cite{Sabourin2009,Ohkura2011}, partially premixed and stratified combustion \cite{Dec2009},
plasma assisted combustion \cite{Starikovskiy2013,Uddi2009,Sun2010}, and cool flames \cite{Won2014}, have been proposed and developed.

However, there are still a number of problems, for example,
(i) for spinning detonation, the influences of the wall curvature and fuel/air mixing on the detonation initiation and propagation modes,
(ii) for high pressure stratified combustion, the ignition to detonation transition at low temperature,
(iii) for plasma assisted combustion, the highly non-equilibrium energy transfer between electrons, electronically and vibrationally excited molecules, and neutral molecules,
(iv) for cool flames, the hydrodynamics, chemical kinetics, and kinetics-transport coupling,
are challenging our current understanding \cite{Ju-Review2014,Ombrello2006,Sun2012,Sun2013}.
All these new combustion concepts involve complicated non-equilibrium chemical and transport processes.

For a long time, the main way people know the combustion process is experimental and theoretical research \cite{Chapman1899,Jouguet1905,Zeldovich1940,Neumann1942,Doering1943,Detonation2000,ZChen_Thesis}. In recent five decades, the numerical simulation of combustion process has achieved great success\cite{ZChen_Thesis,ZChen2015,ZChen2014,ZChen2013,CombustionReview}. To simulate a combustion process, the following steps are needed. (i) Establish a physical model. (ii) Establish discrete control equations. (iii) Numerical experiments and data analysis. Generally speaking, for a combustion system, there are three levels of description which are in the microscopic, mesoscopic and macroscopic scales, respectively. The microscopic scale is generally referred to the description at Molecular Dynamic (MD) level. The main numerical tool is the MD simulation. Via study at this level, the reaction rate equation can be established. The macroscopic scale is generally referred to the description based on Navier-Stokes equations. At this level the mainly concerned are
 Hydrodynamic Non-Equilibrium(HNE) behaviours, specifically, the evolutions of the density, temperature, flow velocity and pressure. The mesoscopic description is generally referred to the description based on the gas kinetic theory, more specifically, the Boltzmann equation. At this level, we can study more details of the interfacial structures and the interplay between the HNE and the Thermodynamic Non-Equilibrium(TNE) behaviours.

What is used in the most engineering applications is the macroscopic/hydrodynamic description. The physical model at this level consists of some specific form of the hydrodynamic equations coupled with some phenomenological reaction rate equation, which is constructed according to the conservation laws of mass, momentum and energy, as well as some suitable simplifications. To establish the discrete control equations, the first step is to choose a coordinate system where the coordinate axes should adapt to the edge of the computational/physical domain. When the computational domain is rectangle, cylindrical or conical, the generally chosen coordinate system can be orthogonal, cylindrical, or spherical. The second step is to establish a structured, unstructured or block-structured grid according to the specific situation. The third step is to choose or formulate a discretization scheme. The frequently used schemes includes the Finite Difference (FD), the finite volume, the finite element, the finite analytic\cite{FA}, the boundary element, the integration transformation, the spectral method, etc.

In recent three decades the Lattice Boltzmann (LB) method \cite%
{Succi1989,Higuera1989,Succi1992,Succi-Book,Succi_RMP,Succi_Science,Yeomans_PRL1997,Yeomans_PRL2001,Yeomans_PRL2002,ShanChen,SYChen,DXZhang,HPFang,Guozhaoli2013%
,ProgPhys2014,Dawson1993,Weimar1996, ZhangRenliang2014,ChenShiyi1996,LaiHuilin} has been becoming a powerful tool to simulate various complex flows.
Due to the importance of combustion phenomena, one can find a number of LB papers in literature \cite{Succi1997,Filippova1998,Filippova2000JCP,Filippova2000CPC,Yu2002%
,Yamamoto2002,Yamamoto2003,Yamamoto2005,Lee2006,Chiavazzo2010%
,ChenSheng2007,ChenSheng2008,ChenSheng2009,ChenSheng2010I,ChenSheng2010II,ChenSheng2010III,ChenSheng2011,ChenSheng2012}. The pioneering LB model for combustion systems was given by Succi et al. \cite{Succi1997} in 1997. This work is based on the assumptions of fast chemistry and cold flames with weak heat release. In the following years, Filippova and H\"{a}nel \cite{Filippova1998,Filippova2000JCP,Filippova2000CPC} proposed a kind of hybrid scheme for low Mach number reactive flows. The flow field is solved by modified lattice-BGK method and the transport equations for energy and species are solved by a FD scheme. Via the LB method Yu et al. \cite{Yu2002} simulated scalar mixing in a multi-component flow and a chemical reacting flow. Yamamoto et al \cite{Yamamoto2002} constructed a LB scheme for combustion phenomena including the reaction, diffusion and convection effects. Lee et al. \cite{Lee2006} presented a Double-Distribution Function LB model to solve the laminar diffusion flames within the context of Burke-Schumann flame sheet model. In recent years Chen et al \cite{ChenSheng2007,ChenSheng2008} developed a coupled LB method for the low Mach number combustion and presented some meaningful results \cite%
{ChenSheng2009,ChenSheng2010I,ChenSheng2010II,ChenSheng2010III,ChenSheng2011,ChenSheng2012}.

In brief, LB modeling of combustion phenomena has long been an interesting topic, but was mainly focused on low Mach number combustion where the incompressible LB model works. In those studies the LB model works as a kind of new scheme to recover the hydrodynamic model. In those thermal LB models, the temperature $T$ could not be described by the same Distribution Function (DF) which describes the density $\rho$ and flow velocity $\mathbf{u}$. In some LB models it was further assumed that the chemical reaction has no effect on the flow field.

As a special case of combustion, the explosion phenomena lead to accidents or disasters sometimes. But the controlled explosion has been widely applied in various engineering problems, such as, explosion painting, explosion cleaning, explosion working, explosion propulsion, demolition blasting, blasting mining, blasting excavation, etc. The traditional Computational Fluid Dynamics (CFD) has been used to simulate explosion for many years. It is interesting to extend the LB model to simulate such complex phenomena. As a special discretization of the Boltzmann equation, the appropriately designed LB model should possess more kinetic information which is  beyond the description of the Navier-Stokes equations. For convenience of description, we refer such a LB kinetic model as to LBKM.

To model a more practical combustion phenomenon, an appropriate LBKM should be thermal, compressible, and work for both the low and the high Mach number flows. At the same time, the chemical reaction and flow behaviour should couple naturally. In such a LBKM, the density $\rho$, flow velocity $\mathbf{u}$, temperature $T$ and relevant higher-order kinetic moments should be described by the same DF. It should work as a new tool to probe both the HNE and TNE \cite{Succi-Book,ProgPhys2014,Review2012}.

In recent years the development of LB models for high speed compressible flows \cite%
{Alexander1992,Alexander1993,Chen1994,McNamara1997,Watari2003%
,XuChenEPL2010,XuGan2013,XuChen2014,XuLin2014PRE} makes it possible to simulate systems with shock wave. Very recently we presented two LBKMs for high Mach combustion and detonation phenomena \cite{XuYan2013,XuLin2014CTP}. The first is in Cartesian coordinates \cite{XuYan2013}. The second \cite{XuLin2014CTP} is in polar coordinate system and designed for simulating the explosion and implosion behaviours. Both the two models are based on the Single-Relaxation-Time (SRT) BGK-Boltzmann equation. Consequently, the Prandtl number is fixed at $1$. A solution to this problem is to use a Multiple-Relaxation-Time (MRT) version of the LBKM. Early in 1989 Higuera,  Succi and Benzi developed a strategy for building suitable collision operators\cite{Succi1989}, which is the precursor of MRT models \cite{Higuera1989,Succi1992,Luo,XuChenEPL2010,XuChen2014}. In this work we present a MRT-LBKM for low and high Mach number combustion phenomena. Besides the viscosity and heat conductivity can be adjusted independently, the model can be used to track the TNE effects and investigate the interplay between the HNE and TNE behaviours.

The rest of the paper is organized as below. In section II the MRT-LBKM for combustion phenomena is formulated. In section III the Chapman-Enskog analysis is performed. The validation and verification of the new model are presented in section IV. Some discussions on the physical gains and computing time of various LB models are shown in section V. In section VI the new model is used to probe some fine structures of the detonation wave. Section VII summarizes and concludes the present paper.

\section{Formulation of the lattice Boltzmann kinetic model}

The practical combustion process is very complicated. To study some fundamental behaviours in the combustion system, in this work we propose a simple LBKM described by the following equation,
\begin{eqnarray}
\frac{\partial f_{i}}{\partial t}+v_{i\alpha }\frac{\partial f_{i}}{\partial
r_{\alpha }} &=&-M_{il}^{-1}\left[ \hat{R}_{lk}\left( \hat{f}_{k}-\hat{f}%
_{k}^{eq}\right) +\hat{A_{l}}\right] +C_{i}\text{,} \label{Boltzmann_chemical}
\\
C_{i} &=&\frac{df_{i}}{dt}|_{C}  \label{chemical_term}
\end{eqnarray}%
where $i$ ($=1$,$2$,$\cdots $,$N$) is the index of discrete velocity, $N$ is
the total number of the discrete velocities used in the LBKM, $f_{i}$ is the
discrete distribution function, $v_{i\alpha }$ is the
$\alpha$- component of the $i$-th discrete
velocity, $\alpha =x$, $y$;
$\hat{f}_{k}=M_{ki}f_{i}$
($\hat{f}_{k}^{eq}=M_{ki}f_{i}^{eq}$)
is the moment of the (equilibrium) distribution function and formally the (equilibrium) distribution function in the moment
space; $M_{ki}$ is the element of the matrix $\mathbf{M}$ connecting the
vector of discrete distribution function, $\mathbf{f}=\left( f_{1}\text{, }%
f_{2}\text{, }\cdots \text{,}f_{N}\right) ^{T}$, and the vector, $\hat{%
\mathbf{f}}=\left( \hat{f}_{1}\text{, }\hat{f}_{2}\text{, }\cdots \text{, }%
\hat{f}_{N}\right) ^{T}$; $\hat{\mathbf{R}}=\mathbf{MRM}^{-1}=$diag$\left(
R_{1},R_{2},\cdots ,R_{N}\right) $ is a diagonal matrix whose element $R_{k}$
describes speed of $\hat{f}_{k}$ approaching $\hat{f}_{k}^{eq}$; $\hat{A}_{l}$ is the $l$%
-th element of $\hat{\mathbf{A}}=(0\text{,}\cdots \text{,}0\text{,}\hat{A}%
_{8}\text{,}\hat{A}_{9}\text{,}0\text{,}\cdots \text{,}0)^{T}$ and is a
modification to the collision operator $\hat{R}_{lk}\left( \hat{f}_{k}-\hat{f%
}_{k}^{eq}\right) $, where
\begin{eqnarray}
\hat{A}_{8} &=&\rho T\frac{R_{5}-R_{8}}{R_{5}}[4u_{x}(\frac{\partial u_{x}}{%
\partial x}-\frac{1}{D+I}\frac{\partial u_{x}}{\partial x}-\frac{1}{D+I}%
\frac{\partial u_{y}}{\partial y})+2u_{y}(\frac{\partial u_{y}}{\partial x}+%
\frac{\partial u_{x}}{\partial y})]\text{,}  \label{ArtificialA8} \\
\hat{A}_{9} &=&\rho T\frac{R_{7}-R_{9}}{R_{7}}[4u_{y}(\frac{\partial u_{y}}{%
\partial y}-\frac{1}{D+I}\frac{\partial u_{x}}{\partial x}-\frac{1}{D+I}%
\frac{\partial u_{y}}{\partial y})+2u_{x}(\frac{\partial u_{y}}{\partial x}+%
\frac{\partial u_{x}}{\partial y})]\text{.}  \label{ArtificialA9}
\end{eqnarray}%

The reason for this modification is as below. Although
from the mathematical point of view, the relaxation coefficient $R_{k}$ can
be independently adjusted for each kinetic mode $\left( \hat{f}_{k}-\hat{f}%
_{k}^{eq}\right) $, from the physical point of view, coupling may exist
between or among different kinetic modes. To ensure the MRT model can
present correct macroscopic behavior, one can perform the Chapman-Enskog
analysis and analyze the consistency of the terms describing transportation
in the recovered hydrodynamic equations to find a solution for the
modification to the collision term\cite {XuChen2014}.
This modification is added so that the LBKM can recover the consistent Navier-Stokes equations in the hydrodynamic limit. $C_{i}$ is the chemical term added to the LB equation and will be given a
specific form in the following part. For convenience of description below,
we introduce $A_{i}=M_{il}^{-1}\hat{A_{l}}$. In this work we consider a
two-dimensional ($D=2$) system where the particle mass is unity. The discrete
equilibrium distribution function satisfies the following relations
\begin{equation}
\sum f_{i}^{eq}=\rho =\sum f_{i}
\text{,} \label{moment1}
\end{equation}
\begin{equation}
\sum f_{i}^{eq}v_{i\alpha }=\rho u_{\alpha }=\sum f_{i}v_{i\alpha }
\text{,} \label{moment2}
\end{equation}
\begin{equation}
\sum f_{i}^{eq}(v_{i}^{2}+\eta _{i}^{2})=\rho [ (D+I)T+u^{2}]=\sum f_{i}(v_{i}^{2}+\eta _{i}^{2})
\text{,} \label{moment3}
\end{equation}
\begin{equation}
\sum f_{i}^{eq}v_{i\alpha }v_{i\beta }=\rho (\delta _{\alpha \beta}T+u_{\alpha }u_{\beta })
\text{,} \label{moment4}
\end{equation}
\begin{equation}
\sum f_{i}^{eq}(v_{i}^{2}+\eta _{i}^{2})v_{i\alpha }=\rho u_{\alpha}[(D+I+2)T+u^{2}]
\text{,} \label{moment5}
\end{equation}
\begin{equation}
\sum f_{i}^{eq}v_{i\alpha }v_{i\beta }v_{i\chi }=\rho (u_{\alpha }\delta
_{\beta \chi }+u_{\beta }\delta _{\chi \alpha }+u_{\chi }\delta _{\alpha
\beta })T+\rho u_{\alpha }u_{\beta }u_{\chi }
\text{,} \label{moment6}
\end{equation}
\begin{equation}
\sum f_{i}^{eq}(v_{i}^{2}+\eta _{i}^{2})v_{i\alpha }v_{i\beta }=\rho \delta
_{\alpha \beta }[(D+I+2)T+u^{2}]T+\rho u_{\alpha }u_{\beta }[(D+I+4)T+u^{2}]%
\text{,} \label{moment7}
\end{equation}
\begin{equation}
\sum f_{i}^{eq}\eta _{i}^{2}v_{i\alpha }v_{i\beta }=\rho \delta _{\alpha
\beta }IT^{2}+\rho u_{\alpha }u_{\beta }IT]
\text{,} \label{moment8}
\end{equation}
\begin{equation}
\sum f_{i}^{eq}(v_{i}^{2}+\eta _{i}^{2})\eta _{i}^{2}=\rho IT[u^{2}+(D+3I)T]%
\text{,}  \label{moment9}
\end{equation}
\begin{equation}
\sum f_{i}^{eq}(v_{i}^{2}+\eta _{i}^{2})v_{i}^{2}v_{i\alpha }=\rho u_{\alpha
}[u^{4}+(D+2)(D+I+4)T^{2}+(2D+I+8)u^{2}T]\text{,}  \label{moment10}
\end{equation}%
\begin{equation}
\sum f_{i}^{eq}(v_{i}^{2}+\eta _{i}^{2})\eta _{i}^{2}v_{i\alpha }=\rho
u_{\alpha }IT[u^{2}+(D+3I+2)T]
\text{,} \label{moment11}
\end{equation}
where $\rho $, $T$, $p$ ($=\rho T$), and $u_{\alpha}$ are the density, temperature, pressure and velocity, respectively. Besides the translational degrees of freedom, $\eta _{i}$ is a free parameter introduced to describe the $I$ extra degrees of freedom corresponding to molecular rotation and/or internal vibration. The internal kinetic energy per unit volume is $E=\rho (D+I)T/2$.

Actually, Eqs.(\ref{moment1})-(\ref{moment11}) can be uniformly written in a matrix form, i.e.,
\begin{equation}
\mathbf{M\times f}^{eq}=\hat{\mathbf{f}}^{eq}, \label{moment12}
\end{equation}
where
the bold-face symbols,
$\mathbf{f}^{eq}=(f_{1}^{eq},f_{2}^{eq},\cdots ,f_{N}^{eq})^{T}$ and $%
\hat{\mathbf{f}}^{eq}=(\hat{f}_{1}^{eq},\hat{f}_{2}^{eq},\cdots ,\hat{f}_{N}^{eq})^{T}$,  denote N-dimensional
column vectors. The matrix $\mathbf{M}=(\mathbf{M}_{1},\mathbf{M}_{2},\cdots ,\mathbf{M}_{N})^{T}$, $%
\mathbf{M}_{i}=(m_{i1},m_{i2},\cdots ,m_{iN})$, where $m_{1i}=1$, $m_{2i}=v_{ix}$, $%
m_{3i}=v_{iy}$, $m_{4i}=v_{i}^{2}+\eta _{i}^{2}$, $m_{5i}=v_{ix}^{2}$, $%
m_{6i}=v_{ix}v_{iy}$, $m_{7i}=$ $v_{iy}^{2}$, $m_{8i}=(v_{i}^{2}+\eta
_{i}^{2})v_{ix}$, $m_{9i}=(v_{i}^{2}+\eta _{i}^{2})v_{iy}$, $%
m_{10i}=v_{ix}^{3}$, $m_{11i}=v_{ix}^{2}v_{iy}$, $m_{12i}=v_{ix}v_{iy}^{2}$,
$m_{13i}=v_{iy}^{3}$, $m_{14i}=(v_{i}^{2}+\eta _{i}^{2})v_{ix}^{2}$, $%
m_{15i}=(v_{i}^{2}+\eta _{i}^{2})v_{ix}v_{iy}$, $m_{16i}=(v_{i}^{2}+\eta
_{i}^{2})v_{iy}^{2}$, $m_{17i}=\eta _{i}^{2}v_{ix}^{2}$, $m_{18i}=\eta
_{i}^{2}v_{ix}v_{iy}$, $m_{19i}=\eta _{i}^{2}v_{iy}^{2}$, $%
m_{20i}=(v_{i}^{2}+\eta _{i}^{2})\eta _{i}^{2}$, $m_{21i}=(v_{i}^{2}+\eta
_{i}^{2})v_{i}^{2}v_{ix}$, $m_{22i}=(v_{i}^{2}+\eta _{i}^{2})v_{i}^{2}v_{iy}$%
, $m_{23i}=(v_{i}^{2}+\eta _{i}^{2})\eta _{i}^{2}v_{ix}$, $%
m_{24i}=(v_{i}^{2}+\eta _{i}^{2})\eta _{i}^{2}v_{iy}$.\ Correspondingly, $%
\hat{f}_{1}^{eq}=\rho $, $\hat{f}_{2}^{eq}=\rho u_{x}$, $\hat{f}%
_{3}^{eq}=\rho u_{y}$, $\hat{f}_{4}^{eq}=\rho [ (D+I)T+u^{2}]$, $%
\hat{f}_{5}^{eq}=\rho (T+u_{x}^{2})$, $\hat{f}_{6}^{eq}=\rho
u_{x}u_{y}$, $\hat{f}_{7}^{eq}=\rho (T+u_{y}^{2})$, $\hat{f}%
_{8}^{eq}=\rho u_{x}[(D+I+2)T+u^{2}]$, $\hat{f}_{9}^{eq}=\rho
u_{y}[(D+I+2)T+u^{2}]$, $\hat{f}_{10}^{eq}=3\rho u_{x}T+\rho u_{x}^{3}$,
$\hat{f}_{11}^{eq}=\rho u_{y}T+\rho u_{x}^{2}u_{y}$, $\hat{f}%
_{12}^{eq}=\rho u_{x}T+\rho u_{x}u_{y}^{2}$, $\hat{f}_{13}^{eq}=3\rho
u_{y}T+\rho u_{y}^{3}$, $\hat{f}_{14}^{eq}=\rho [ (D+I+2)T+u^{2}]+\rho
u_{x}^{2}[(D+I+4)T+u^{2}]$, $\hat{f}_{15}^{eq}=\rho
u_{x}u_{y}[(D+I+4)T+u^{2}]$, $\hat{f}_{16}^{eq}=\rho [
(D+I+2)T+u^{2}]+\rho u_{y}^{2}[(D+I+4)T+u^{2}]$, $\hat{f}_{17}^{eq}=\rho
IT^{2}+\rho u_{x}^{2}IT$, $\hat{f}_{18}^{eq}=\rho u_{x}u_{y}IT$, $%
\hat{f}_{19}^{eq}=\rho IT^{2}+\rho u_{y}^{2}IT$, $\hat{f}%
_{20}^{eq}=\rho IT[u^{2}+(D+3I)T]$, $\hat{f}_{21}^{eq}=\rho
u_{x}[u^{4}+(D+2)(D+I+4)T^{2}+(2D+I+8)u^{2}T]$, $\hat{f}_{22}^{eq}=\rho
u_{y}[u^{4}+(D+2)(D+I+4)T^{2}+(2D+I+8)u^{2}T]$, $\hat{f}_{23}^{eq}=\rho
u_{x}IT[u^{2}+(D+3I+2)T]$, $\hat{f}_{24}^{eq}=\rho
u_{y}IT[u^{2}+(D+3I+2)T]$.

Formally, compared with the MRT version of the LBKM for high speed compressible
flows \cite{XuChenEPL2010,XuChen2014}, the second term, $C_{i}$, in the right sides of
Eq.(\ref{Boltzmann_chemical}) describes the variation of distribution function due to the chemical reaction.

The actual combustion processes are very complicated. In this work we consider the simple combustion processes  and present a simple LBKM based on the following assumptions:

1. The flow behaviour is described by a single distribution function $f$. The the relaxation coefficient $R_{k}$ is a constant, where $k=1$, $2$, $\cdots $, $N$.

2. The radiative heat loss is neglected.

3. The reaction process is irreversible and described by an empirical or semi-empirical equation,
\begin{equation}
\lambda ^{\prime }=\frac{d\lambda }{dt}=F(\lambda ) \text{,}
\label{lambda_formula}
\end{equation}
where $\lambda =\rho _{p}/\rho $ is the concentration of the reaction product in the system and denotes the progress of the reaction; $\rho _{p}$ is the density of the reaction product;  $\rho$ is the density of the whole system.

4. The chemical energy is directly transformed into the internal energy in the following form
\begin{equation}
\frac{dE}{dt}|_{C}=\rho Q \lambda ^{\prime }
\label{chemical_energy}
\end{equation}
where $Q$ is the amount of heat released by the chemical reactant per unit mass.

5. The chemical reaction is slow enough, compared with kinetic process of
approaching thermodynamic equilibrium, so that
\begin{equation}
\frac{df}{dt}|_{C}\approx \frac{df^{eq}}{dt}|_{C} \text{.}
\label{M_chemical_1}
\end{equation}

The chemical reaction results only in the increase of local temperature $T$ and
the local density $\rho $ and hydrodynamic velocity $\mathbf{u}$ remain unchanged. Thus,
\begin{equation}
\frac{df^{eq}}{dt}|_{C}=\frac{\partial f^{eq}}{\partial T}\frac{dT}{dt}|_{C}
\text{.} \label{M_chemical_2}
\end{equation}
The equilibrium distribution function $f^{eq}$ used here reads
\begin{equation}
f^{eq}=\rho (\frac{1}{2\pi T})^{D/2}(\frac{1}{2\pi IT})^{1/2}Exp[-\frac{(%
\mathbf{v}-\mathbf{u})^{2}}{2T}-\frac{\eta ^{2}}{2IT}]
 \label{feq_continuous}
\end{equation}
which gives
\begin{equation}
\frac{\partial f^{eq}}{\partial T}=
\frac{-(1+D)IT+I(\mathbf{v}-\mathbf{u})^{2}+\eta ^{2}}{2IT^{2}}f^{eq}
\text{.} \label{M_chemical_feq}
\end{equation}
It is easy to get
\begin{equation}
\frac{dT}{dt}|_{C}=\frac{2Q}{D+I}F(\lambda )\text{.}
\label{M_chemical_T}
\end{equation}
from the relation $E=\rho (D+I) T/2$ and Eqs.(\ref{lambda_formula})-(\ref{chemical_energy}).
Substituting Eqs.(\ref{M_chemical_feq})-(\ref{M_chemical_T}) into (\ref{M_chemical_2}) gives
\begin{equation}
C_{i}=f_{i}^{eq}Q\frac{-(1+D)IT+I(\mathbf{v}_{i}-\mathbf{u})^{2}
+\eta_{i}^{2}}{I(D+I)T^{2}}F(\lambda )
\text{.} \label{C_i}
\end{equation}

\begin{figure}[tbp]
\center\includegraphics*
[bbllx=0pt,bblly=0pt,bburx=205pt,bbury=180pt,angle=0,width=0.4\textwidth]{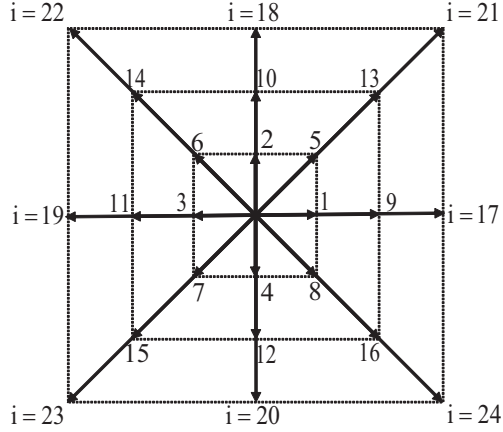}
\caption{Schematic of the discrete velocity model.}
\label{Fig01}
\end{figure}
Equations (\ref{Boltzmann_chemical}) can be rewritten as twenty-four coupled equations in the two-dimensional case. Consequently we need a Discrete Velocity Model (DVM) with at least $24$ discrete velocities. To obtain the high computational efficiency, we choose the following two-dimensional DVM which only has $24$ discrete velocities (see Fig.\ref{Fig01}),
\begin{eqnarray}
\overline{\mathbf{v}}_{i} &=&\left\{
\begin{array}{cccc}
cyc: & (\pm 1,0) & for & 1\leq i\leq 4\text{,} \\
cyc: & (\pm 1,\pm 1) & for & 5\leq i\leq 8\text{,}%
\end{array}%
\right. \\
\mathbf{v}_{i} &=&\left\{
\begin{array}{ccc}
v_{a}\overline{\mathbf{v}}_{i} & for & 1\leq i\leq 8\text{,} \\
v_{b}\overline{\mathbf{v}}_{i-8} & for & 8\leq i\leq 16\text{,} \\
v_{c}\overline{\mathbf{v}}_{i-16} & for & 17\leq i\leq 24\text{,}%
\end{array}%
\right. \\
\eta_{i} &=&\left\{
\begin{array}{ccc}
\eta_{a} & for & 1\leq i\leq 8\text{,} \\
\eta_{b} & for & 8\leq i\leq 16\text{,} \\
\eta_{c} & for & 17\leq i\leq 24\text{,}%
\end{array}%
\right.
\end{eqnarray}%
where \textbf{cyc} indicates the cyclic permutation. For convenience of description, we refer the two-dimensional DVM with $24$ discrete velocities as to D2V24.

It has been known that the spurious oscillations occuring near shock wave with finite-difference equations are related to the dispersion term in the corresponding modified differential equations. If the sign of the dispersion coefficient, say $\nu$, is properly adjusted, that is, the sign changes across shock wave, $\nu >0$ upstream, and $\nu <0$  downstream, the undesirable oscillations can be totally suppressed.
Therefore, in this work, the spatial derivatives in Eq.(\ref{Boltzmann_chemical}) are calculated by adopting the Nonoscillatory and Nonfree-parameters Dissipative (NND) finite difference scheme \cite{Zhang1988}.
The evolution of chemical process is described by
\begin{equation}
\frac{\partial \lambda }{\partial t}+u_{\alpha }\frac{\partial \lambda }{%
\partial r_{\alpha }}=\omega _{1}p^{m}(1-\lambda )+\omega _{2}p^{n}\lambda
(1-\lambda )\text{,}  \label{lambda_equation}
\end{equation}
where the so-called Cochran's rate function \cite{Cochran1979} is adopted for the
description of chemical reaction; $\omega _{1}$, $\omega _{2}$, $m$ and $n$ are adjustable parameters. Without lossing generality, the ignition temperature $T_{ig}=1.1$ is assumed in this work. Only when $T>T_{ig}$ can the chemical reaction proceed. And we choose the parameters, $m=n=1$. The temporal derivative in Eq.(\ref{lambda_equation}) is solved analytically, and the spatial ones by the NND
scheme \cite{Zhang1988}.

The inverse of the matrix $\mathbf{M}$ can be analytically solved by the software, Matlab2011.
It should be pointed out that, although the complete formulas seems long and complicated, in practical simulations, the parameters ($v_{a}$, $v_{b}$, $v_{c}$, $\eta_{a}$, $\eta_{b}$, $\eta_{c}$) are replaced by specific values, then the elements of matrix $\mathbf{M}$ and its inverse $\mathbf{M}^{-1}$ are fixed also by specific values before the main loop.

\section{Chapman-Enskog analysis of the model}

The Chapman-Enskog analysis shows that, only if $\mathbf{f}^{eq}$ satisfies the statistical relation, (\ref{moment12}), or specifically,
the eleven equations, (\ref{moment1})-(\ref{moment11}), then the LB equation (\ref{Boltzmann_chemical}) can recover the Navier-Stokes model for combustion. We show the main procedure of the Chapman-Enskog analysis  below.

From Eq.(\ref{Boltzmann_chemical}), we get
\begin{equation}
\frac{\partial \hat{\mathbf{f}}}{\partial t}+\frac{\partial }{\partial
r_{\alpha }}(\hat{\mathbf{E}}_{\alpha }\hat{\mathbf{f}})=-\hat{\mathbf{R}}\left( \hat{\mathbf{f}}-\hat{\mathbf{f}}^{eq}\right)+\hat{\mathbf{A}}+\hat{\mathbf{C}}
\text{,} \label{CE_LB}
\end{equation}
where $\hat{\mathbf{C}}=\mathbf{MC}$, $\hat{\mathbf{E}}_{\alpha }=%
\mathbf{Mv}_{\alpha }\mathbf{M}^{-1}$, and $\mathbf{v}_{\alpha }=$diag$%
(v_{1\alpha },v_{2\alpha },\cdots ,v_{N\alpha })$ is a diagonal matrix.

Expanding the variables with respect to $\varepsilon$ corresponding the Knudsen number, as
\begin{equation}
\left\{
\begin{array}{lll}
f_{i}&=&f_{i}^{(0)}+f_{i}^{(1)}+f_{i}^{(2)}+\cdots \\
A_{i}&=&A_{1i} \\
C_{i}&=&C_{1i} \\
\frac{\partial }{\partial t}&=&\frac{\partial }{\partial t_{1}}+\frac{\partial}{\partial t_{2}} \\
\frac{\partial }{\partial r_{\alpha }}&=&\frac{\partial }{\partial r_{1\alpha }}
\end{array}
\right.  \label{CE_expansion}
\end{equation}
where the part of distribution function $f_{i}^{(l)}=\mathrm{O}(\varepsilon ^{l})$, the modification term $A_{1i}=\mathrm{O}(\varepsilon)$, the chemical term $C_{1i}=\mathrm{O}(\varepsilon)$, the partial derivatives $\partial /\partial t_{l}=\mathrm{O}(\varepsilon ^{l})$ and $\partial /\partial r_{1 \alpha}=\mathrm{O}(\varepsilon)$, ($l=1,2,\cdots $). It is easy to get from the first three subequations of Eq.(\ref{CE_expansion}) that
\begin{equation}
\hat{f}_{i}=\hat{f}_{i}^{(0)}+\hat{f}_{i}^{(1)}+\hat{f}_{i}^{(2)}+\cdots \text{,}  \label{CE_fcap_expansion}
\end{equation}
\begin{equation}
\hat{A}_{i}=\hat{A}_{1i}\text{,}  \label{CE_Acap_expansion}
\end{equation}
\begin{equation}
\hat{C}_{i}=\hat{C}_{1i}\text{.}  \label{CE_Ccap_expansion}
\end{equation}
By substituting the last two subequations of (\ref{CE_expansion}) and Eqs.(\ref{CE_fcap_expansion})-(\ref{CE_Ccap_expansion}) into (\ref{CE_LB}) and comparing
the coefficients of the same order of $\varepsilon $, we have
\begin{equation}
\mathrm{O}(\varepsilon ^{0}):\ \ \hat{\mathbf{f}}^{(0)}=\hat{\mathbf{f}}^{eq},  \label{CE_epsilon0}
\end{equation}
\begin{equation}
\mathrm{O}(\varepsilon ^{1})\text{: \ }(\frac{\partial }{\partial t_{1}}+%
\hat{\mathbf{E}}_{\alpha }\frac{\partial }{\partial r_{1\alpha }})%
\hat{\mathbf{f}}^{(0)}=-\hat{\mathbf{R}}\hat{\mathbf{f}}^{(1)}
+\hat{\mathbf{A}}+\hat{\mathbf{C}}\text{,}  \label{CE_epsilon1}
\end{equation}
\begin{equation}
\mathrm{O}(\varepsilon ^{2})\text{: \ }\frac{\partial }{\partial t_{2}}%
\hat{\mathbf{f}}^{(0)}+(\frac{\partial }{\partial t_{1}}+\hat{%
\mathbf{E}}_{\alpha }\frac{\partial }{\partial r_{1\alpha }})\hat{%
\mathbf{f}}^{(1)}=-\hat{\mathbf{R}}\hat{\mathbf{f}}^{(2)}\text{,}
\label{CE_epsilon2}
\end{equation}
where $\mathbf{f}^{(l)}=\left( f_{1}^{(l)}\text{, }f_{2}^{(l)}\text{, }%
\cdots \text{,}f_{N}^{(l)}\right) ^{T}$. Specifically, $\mathbf{f}^{(0)}$\ is
the matrix for the equilibria of the moments, $\mathbf{f}%
^{(1)}$\ and $\mathbf{f}^{(2)}$ are the matrixes for the first order and
second order deviations from equilibria.

From Eq.(\ref{CE_epsilon1}), we get
\begin{equation}
\frac{\partial \hat{f}_{{1}}^{eq}}{\partial t_{1}}+\frac{\partial
\hat{f}_{{2}}^{eq}}{\partial x_{1}}+\frac{\partial \hat{f}_{{3}}^{eq}%
}{\partial y_{1}}=-R_{1}\hat{f}_{1}^{(1)}+\hat{C}_{1}\text{,}
\label{CE_f1caq1}
\end{equation}
\begin{equation}
\frac{\partial \hat{f}_{{2}}^{eq}}{\partial t_{1}}+\frac{\partial
\hat{f}_{{5}}^{eq}}{\partial x_{1}}+\frac{\partial \hat{f}_{{6}}^{eq}%
}{\partial y_{1}}=-R_{2}\hat{f}_{2}^{(1)}+\hat{C}_{2}\text{,}
\label{CE_f1caq2}
\end{equation}
\begin{equation}
\frac{\partial \hat{f}_{{3}}^{eq}}{\partial t_{1}}+\frac{\partial
\hat{f}_{{6}}^{eq}}{\partial x_{1}}+\frac{\partial \hat{f}_{{7}}^{eq}%
}{\partial y_{1}}=-R_{3}\hat{f}_{3}^{(1)}+\hat{C}_{3}\text{,}
\label{CE_f1caq3}
\end{equation}
\begin{equation}
\frac{\partial \hat{f}_{{4}}^{eq}}{\partial t_{1}}+\frac{\partial
\hat{f}_{{8}}^{eq}}{\partial x_{1}}+\frac{\partial \hat{f}_{{9}}^{eq}%
}{\partial y_{1}}=-R_{3}\hat{f}_{4}^{(1)}+\hat{C}_{4}\text{,}
\label{CE_f1caq4}
\end{equation}
\begin{equation}
\frac{\partial \hat{f}_{{5}}^{eq}}{\partial t_{1}}+\frac{\partial
\hat{f}_{{10}}^{eq}}{\partial x_{1}}+\frac{\partial \hat{f}_{{11}%
}^{eq}}{\partial y_{1}}=-R_{5}\hat{f}_{5}^{(1)}+\hat{C}_{5}\text{,}
\label{CE_f1caq5}
\end{equation}
\begin{equation}
\frac{\partial \hat{f}_{{6}}^{eq}}{\partial t_{1}}+\frac{\partial
\hat{f}_{{11}}^{eq}}{\partial x_{1}}+\frac{\partial \hat{f}_{{12}%
}^{eq}}{\partial y_{1}}=-R_{6}\hat{f}_{6}^{(1)}+\hat{C}_{6}\text{,}
\label{CE_f1caq6}
\end{equation}
\begin{equation}
\frac{\partial \hat{f}_{{7}}^{eq}}{\partial t_{1}}+\frac{\partial
\hat{f}_{{12}}^{eq}}{\partial x_{1}}+\frac{\partial \hat{f}_{{13}%
}^{eq}}{\partial y_{1}}=-R_{7}\hat{f}_{7}^{(1)}+\hat{C}_{7}\text{,}
\label{CE_f1caq7}
\end{equation}
\begin{equation}
\frac{\partial \hat{f}_{{8}}^{eq}}{\partial t_{1}}+\frac{\partial
\hat{f}_{{14}}^{eq}}{\partial x_{1}}+\frac{\partial \hat{f}_{{15}%
}^{eq}}{\partial y_{1}}=-R_{8}\hat{f}_{8}^{(1)}
+\hat{A}_{8}+\hat{C}_{8}\text{,}
\label{CE_f1caq8}
\end{equation}
\begin{equation}
\frac{\partial \hat{f}_{{9}}^{eq}}{\partial t_{1}}+\frac{\partial
\hat{f}_{{15}}^{eq}}{\partial x_{1}}+\frac{\partial \hat{f}_{{16}%
}^{eq}}{\partial y_{1}}=-R_{9}\hat{f}_{9}^{(1)}
+\hat{A}_{9}+\hat{C}_{9}\text{,}
\label{CE_f1caq9}
\end{equation}

It is easy to get from Eqs. (\ref{moment1})-(\ref{moment11}) and (\ref{C_i}) that $\hat{C}_{1}=0$, $\hat{C}_{2}=0$, $\hat{C}_{3}=0 $, $\hat{C}_{4}=2\rho \lambda ^{\prime }Q$, $\hat{C}_{5}=2\rho \lambda ^{\prime }Q/(D+I)$, $\hat{C}_{6}=0$, $\hat{C}_{7}=2\rho \lambda ^{\prime }Q/(D+I)$, $\hat{C}_{8}=2\rho u_{x}\lambda ^{\prime }Q (D+I+2)/(D+I)$, $\hat{C}_{9}=2\rho u_{y}\lambda ^{\prime }Q (D+I+2)/(D+I)$. Substituting the all the specific
forms of $\hat{C}_{i}$ and $\hat{f}_{i}^{eq}$ into (\ref{CE_f1caq1})-(\ref{CE_f1caq9}) gives
\begin{equation}
\frac{\partial \rho }{\partial t_{1}}+\frac{\partial j_{x}}{\partial x_{1}}+%
\frac{\partial j_{y}}{\partial y_{1}}=0\text{,}  \label{CE_order1_1}
\end{equation}%
\begin{equation}
\frac{\partial j_{x}}{\partial t_{1}}+\frac{\partial \rho (T+u_{x}^{2})}{%
\partial x_{1}}+\frac{\partial \rho u_{x}u_{y}}{\partial y_{1}}=0\text{,}
\label{CE_order1_2}
\end{equation}%
\begin{equation}
\frac{\partial j_{y}}{\partial t_{1}}+\frac{\partial \rho u_{x}u_{y}}{%
\partial x_{1}}+\frac{\partial \rho (T+u_{y}^{2})}{\partial y_{1}}=0\text{,}
\label{CE_order1_3}
\end{equation}%
\begin{equation}
\frac{\partial \xi }{\partial t_{1}}+\frac{\partial \rho
u_{x}[(D+I+2)T+u^{2}]}{\partial x_{1}}+\frac{\partial \rho
u_{y}[(D+I+2)T+u^{2}]}{\partial y_{1}}=2\rho \lambda ^{\prime }Q\text{,}
\label{CE_order1_4}
\end{equation}%
\begin{equation}
\frac{\partial \rho (T+u_{x}^{2})}{\partial t_{1}}+\frac{\partial \rho
(3u_{x}T+u_{x}^{3})}{\partial x_{1}}+\frac{\partial \rho
(u_{y}T+u_{x}^{2}u_{y})}{\partial y_{1}}=-R_{5}\hat{f}_{5}^{(1)}+\rho
\lambda ^{\prime }Q\frac{2}{D+I}\text{,}  \label{CE_order1_5}
\end{equation}%
\begin{equation}
\frac{\partial \rho u_{x}u_{y}}{\partial t_{1}}+\frac{\partial \rho
(u_{y}T+u_{x}^{2}u_{y})}{\partial x_{1}}+\frac{\partial \rho
(u_{x}T+u_{x}u_{y}^{2})}{\partial y_{1}}=-R_{6}\hat{f}_{6}^{(1)}\text{,}
\label{CE_order1_6}
\end{equation}%
\begin{equation}
\frac{\partial \rho (T+u_{y}^{2})}{\partial t_{1}}+\frac{\partial \rho
(u_{x}T+u_{x}u_{y}^{2})}{\partial x_{1}}+\frac{\partial \rho
(3u_{y}+u_{y}^{3})}{\partial y_{1}}=-R_{7}\hat{f}_{7}^{(1)}+\frac{2\rho
\lambda ^{\prime }Q}{D+I}\text{,}  \label{CE_order1_7}
\end{equation}
\begin{gather}
\frac{\partial \rho u_{x}[(D+I+2)T+u^{2}]}{\partial t_{1}}+\frac{\partial }{%
\partial x_{1}}\{\rho \lbrack (D+I+2)T+u^{2}]T+\rho
u_{x}^{2}[(D+I+4)T+u^{2}]\}  \notag \\
+\frac{\partial \rho u_{x}u_{y}[(D+I+4)T+u^{2}]}{\partial y_{1}}
=-R_{8}\hat{f}_{8}^{(1)}
+\hat{A}_{8}
+2\rho u_{x}\lambda ^{\prime }Q\frac{D+I+2}{D+I}\text{,}
\label{CE_order1_8}
\end{gather}
\begin{gather}
\frac{\partial \rho u_{y}[(D+I+2)T+u^{2}]}{\partial t_{1}}+\frac{\partial }{%
\partial y_{1}}\{\rho \lbrack (D+I+2)T+u^{2}]T+\rho
u_{y}^{2}[(D+I+4)T+u^{2}]\}  \notag \\
+\frac{\partial \rho u_{x}u_{y}[(D+I+4)T+u^{2}]}{\partial x_{1}}
=-R_{9}\hat{f}_{9}^{(1)}
+\hat{A}_{9}
+2\rho u_{y}\lambda ^{\prime }Q\frac{D+I+2}{D+I}\text{,}
\label{CE_order1_9}
\end{gather}
where $j_{x}=\rho u_{x}$, $j_{y}=\rho u_{y}$, and $\xi =(D+I)\rho
T+(j_{x}^{2}+j_{y}^{2})/\rho $ is twice the total energy.

From Eq.(\ref{CE_epsilon2}), we get
\begin{equation}
\frac{\partial \rho }{\partial t_{2}}=0\text{,}  \label{CE_order2_1}
\end{equation}%
\begin{equation}
\frac{\partial j_{x}}{\partial t_{2}}+\frac{\partial \hat{f}_{5}^{(1)}}{%
\partial x_{1}}+\frac{\partial \hat{f}_{6}^{(1)}}{\partial y_{1}}=0\text{%
,}  \label{CE_order2_2}
\end{equation}%
\begin{equation}
\frac{\partial j_{y}}{\partial t_{2}}+\frac{\partial \hat{f}_{6}^{(1)}}{%
\partial x_{1}}+\frac{\partial \hat{f}_{7}^{(1)}}{\partial y_{1}}=0\text{%
,}  \label{CE_order2_3}
\end{equation}%
\begin{equation}
\frac{\partial \xi }{\partial t_{2}}+\frac{\partial \hat{f}_{8}^{(1)}}{%
\partial x_{1}}+\frac{\partial \hat{f}_{9}^{(1)}}{\partial y_{1}}=0\text{%
.}  \label{CE_order2_4}
\end{equation}%
Adding Eqs.(\ref{CE_order1_1})-(\ref{CE_order1_4}) and (\ref{CE_order2_1})-(\ref%
{CE_order2_4}) leads to the following equations,
\begin{equation}
\frac{\partial \rho }{\partial t}+\frac{\partial j_{x}}{\partial x}+\frac{%
\partial j_{y}}{\partial y}=0
\text{,}  \label{NS_1}
\end{equation}
\begin{equation}
\frac{\partial j_{x}}{\partial t}+\frac{\partial \rho (T+u_{x}^{2})}{%
\partial x}+\frac{\partial \rho u_{x}u_{y}}{\partial y}+\frac{\partial
\hat{f}_{5}^{(1)}}{\partial x}+\frac{\partial \hat{f}_{6}^{(1)}}{%
\partial y}=0
\text{,}  \label{NS_2}
\end{equation}
\begin{equation}
\frac{\partial j_{y}}{\partial t}+\frac{\partial \rho u_{x}u_{y}}{\partial x}%
+\frac{\partial \rho (T+u_{y}^{2})}{\partial y}+\frac{\partial \hat{f}%
_{6}^{(1)}}{\partial x}+\frac{\partial \hat{f}_{7}^{(1)}}{\partial y}=0%
\text{,}  \label{NS_3}
\end{equation}
\begin{equation}
\frac{\partial \xi }{\partial t}+\frac{\partial \rho u_{x}[(D+I+2)T+u^{2}]}{%
\partial x}+\frac{\partial \rho u_{y}[(D+I+2)T+u^{2}]}{\partial y}+\frac{%
\partial \hat{f}_{8}^{(1)}}{\partial x}+\frac{\partial \hat{f}%
_{9}^{(1)}}{\partial y}=2\rho \lambda ^{\prime }Q
\text{,}  \label{NS_4}
\end{equation}
From Eqs.(\ref{CE_order1_5})-(\ref{CE_order1_9}) and (\ref{NS_1})-(\ref{NS_4}), we finally obtain the Navier-Stokes  equations:
\begin{equation}
\frac{\partial \rho }{\partial t}+\frac{\partial j_{x}}{\partial x}+\frac{%
\partial j_{y}}{\partial y}=0\text{,}
\end{equation}
\begin{gather}
\frac{\partial j_{x}}{\partial t}+\frac{\partial (p+\rho u_{x}^{2})}{%
\partial x}+\frac{\partial \rho u_{x}u_{y}}{\partial y}  \notag \\
=\frac{\partial }{\partial x}[\frac{\rho T}{R_{5}}(2\frac{\partial u_{x}}{%
\partial x}-\frac{2}{D+I}\frac{\partial u_{x}}{\partial x}-\frac{2}{D+I}%
\frac{\partial u_{y}}{\partial y})]+\frac{\partial }{\partial y}[\frac{\rho T%
}{R_{6}}(\frac{\partial u_{x}}{\partial y}+\frac{\partial u_{y}}{\partial x}%
)]\text{,}
\end{gather}
\begin{gather}
\frac{\partial j_{y}}{\partial t}+\frac{\partial \rho u_{x}u_{y}}{\partial x}%
+\frac{\partial (p+\rho u_{y}^{2})}{\partial y}  \notag \\
=\frac{\partial }{\partial x}[\frac{\rho T}{R_{6}}(\frac{\partial u_{x}}{%
\partial y}+\frac{\partial u_{y}}{\partial x})]+\frac{\partial }{\partial y}[%
\frac{\rho T}{R_{7}}(2\frac{\partial u_{y}}{\partial y}-\frac{2}{D+I}\frac{%
\partial u_{x}}{\partial x}-\frac{2}{D+I}\frac{\partial u_{y}}{\partial y})]%
\label{NSMoment} \text{,}
\end{gather}
\begin{gather}
\frac{\partial \xi }{\partial t}+\frac{\partial (\xi +2p)u_{x}}{\partial x}+%
\frac{\partial (\xi +2p)u_{y}}{\partial y} \notag \\
=2\frac{\partial }{\partial x}[\frac{\rho T}{R_{8}}(c_{p}\frac{\partial T}{%
\partial x}-\frac{2u_{x}}{D+I}\frac{\partial u_{x}}{\partial x}-\frac{2u_{x}%
}{D+I}\frac{\partial u_{y}}{\partial y}+2u_{x}\frac{\partial u_{x}}{\partial
x}+u_{y}\frac{\partial u_{x}}{\partial y}+u_{y}\frac{\partial u_{y}}{%
\partial x})-\frac{1}{2}\frac{\hat{A}_{8}}{R_{8}}] \notag \\
+2\frac{\partial }{\partial y}[\frac{\rho T}{R_{9}}(c_{p}\frac{\partial T}{%
\partial y}-\frac{2u_{y}}{D+I}\frac{\partial u_{x}}{\partial x}-\frac{2u_{y}%
}{D+I}\frac{\partial u_{y}}{\partial y}+2u_{y}\frac{\partial u_{y}}{\partial
y}+u_{x}\frac{\partial u_{y}}{\partial x}+u_{x}\frac{\partial u_{x}}{%
\partial y})-\frac{1}{2}\frac{\hat{A}_{9}}{R_{9}}] \notag \\
+2\rho \lambda ^{\prime }Q
\label{NSEnergy_old} \text{,}
\end{gather}%
Here $c_{p}=(D+I+2)/2$ is the specific-heat at constant pressure. The
specific-heat at constant volume can be defined as $c_{v}=(D+I)/2$.
Substituting Eqs. (\ref{ArtificialA8}) and (\ref{ArtificialA9}) into the
above equation (\ref{NSEnergy_old}) gives
\begin{gather}
\frac{\partial \xi }{\partial t}+\frac{\partial (\xi +2p)u_{x}}{\partial x}+%
\frac{\partial (\xi +2p)u_{y}}{\partial y} \notag \\
=2\frac{\partial }{\partial x}[c_{p}\frac{\rho T}{R_{8}}\frac{\partial T}{%
\partial x}+\frac{\rho T}{R_{5}}(-\frac{2u_{x}}{D+I}\frac{\partial u_{x}}{%
\partial x}-\frac{2u_{x}}{D+I}\frac{\partial u_{y}}{\partial y}+2u_{x}\frac{%
\partial u_{x}}{\partial x}+u_{y}\frac{\partial u_{x}}{\partial y}+u_{y}%
\frac{\partial u_{y}}{\partial x})] \notag \\
+2\frac{\partial }{\partial y}[c_{p}\frac{\rho T}{R_{9}}\frac{\partial T}{%
\partial y}+\frac{\rho T}{R_{7}}(-\frac{2u_{y}}{D+I}\frac{\partial u_{x}}{%
\partial x}-\frac{2u_{y}}{D+I}\frac{\partial u_{y}}{\partial y}+2u_{y}\frac{%
\partial u_{y}}{\partial y}+u_{x}\frac{\partial u_{y}}{\partial x}+u_{x}%
\frac{\partial u_{x}}{\partial y})] \notag \\
+2\rho \lambda ^{\prime }Q
\label{NSEnergy_new} \text{,}
\end{gather}
It is clear to find that, by substituting the specific form of $\hat{A}_{8}$ and $\hat{A}_{9}$ into Eq.(\ref{NSEnergy_old}), the viscous coefficient in the energy equation (\ref{NSEnergy_new}) is consistent with that in the momentum equation (\ref{NSMoment}).
Up to this step, we can find that the proposed MRT-LBKM recover the consistent Navier-Stokes equations in the hydrodynamic limit.

More discussions are as below.
The coefficient $\hat{\mathbf{R}}$ represents
 the inverse of the relaxation time from $\hat{\mathbf{f}}$ to its
equilibrium $\mathbf{\hat{f}^{eq}}$. $\hat{f}%
_{1}=\hat{f}_{1}^{eq}$, $\hat{f}_{2}=\hat{f}_{2}^{eq}$, $\hat{f}_{3}=\hat{f}%
_{3}^{eq}$, $\hat{f}_{4}=\hat{f}_{4}^{eq}$. Consequently, the values of $%
R_{1}$, $R_{2}$, $R_{3}$, $R_{4}$ have no influence on the LB evolution.
Furthermore, the relaxation parameters $R_{i}$ are not completely independent for the system with isotropy constraints \cite{XuChenEPL2010}. Specifically, $R_{5}$, $R_{6}$, $R_{7}$ are related to viscosity, and the viscosity coefficient is $\mu =\rho T/R_{\mu }$ when $%
R_{5}=R_{6}=R_{7}=R_{\mu }$; $R_{8}$, $R_{9}$ are related to heat
conductivity, and the heat conductivity coefficient is $\kappa =c_{p}\rho
T/R_{\kappa }$ when $R_{8}=R_{9}=R_{\kappa }$.
Consequently, both the specific-heat ratio,
\begin{equation}
\gamma =\frac{c_{p}}{c_{v}}=\frac{D+I+2}{D+I}\text{,}
\end{equation}
and the Prandtl number,
\begin{equation}
Pr=\frac{c_{p}\mu }{\kappa }=\frac{R_{\kappa }}{R_{\mu }}\text{,}
\end{equation}
are flexible in this model. When $R_{5}=R_{6}=R_{7}=R_{\mu }$, $R_{8}=R_{9}=R_{\kappa }$,
the above Navier-Stokes equations reduce to
\begin{equation}
\frac{\partial \rho }{\partial t}+\frac{\partial j_{\alpha }}{\partial
r_{\alpha }}=0\text{,}
\end{equation}
\begin{equation}
\frac{\partial j_{\alpha }}{\partial t}+\frac{\partial p}{\partial r_{\alpha
}}+\frac{\partial \rho u_{\alpha }u_{\beta }}{\partial r_{\beta }}=-\frac{%
\partial P_{\alpha \beta }}{\partial r_{\beta }}\text{,}
\end{equation}
\begin{equation}
\frac{\partial \xi }{\partial t}+\frac{\partial (\xi +2p)u_{\alpha }}{%
\partial r_{\alpha }}=2\rho \lambda ^{\prime }Q+2\frac{\partial }{\partial
r_{\beta }}(\kappa \frac{\partial T}{\partial r_{\beta }}-P_{\alpha \beta
}u_{\alpha })
\end{equation}
where
\begin{equation}
\mu_{B}=\mu(\frac{2}{3}-\frac{2}{D+I})
\end{equation}
\begin{equation}
P_{\alpha \beta }=-\mu (\frac{\partial u_{\alpha }}{\partial r_{\beta }}+%
\frac{\partial u_{\beta }}{\partial r_{\alpha }}-\frac{2}{3}\frac{\partial
u_{\chi }}{\partial r_{\chi }}\delta _{\alpha \beta })-\mu _{B}\frac{%
\partial u_{\chi }}{\partial r_{\chi }}\delta _{\alpha \beta }.
\end{equation}
Specifically, $P_{xx}=\hat{f}_{5}^{(1)}$, $P_{xy}=P_{yx}=\hat{f}_{6}^{(1)}$, $P_{yy}=\hat{f}_{7}^{(1)}$.

\section{Numerical tests of the model}

To validate and verify the newly proposed LBKM, here we show simulation results of some well-known benchmark numerical examples which include one for the steady detonation, three for the Riemann problems, one for the shock reflection and one for the Couette flow. The parameter for chemical reaction heat $Q$ is not zero only for the first numerical test. For the cases with the Couette flow, results with different specific heat ratios and Prandtl numbers are shown.

\subsection{Steady detonation}

As the first numerical test, a one-dimensional steady detonation is simulated here to validate our model. The initial physical quantities are:
\begin{equation}
\left\{
\begin{array}{l}
(\rho ,T,u_{x},u_{y},\lambda )_{L}=(1.38837,1.57856,0.577350,0,1) \\
(\rho ,T,u_{x},u_{y},\lambda )_{R}=(1,1,0,0,0)
\end{array}
\right.  \label{V_and_V}
\end{equation}
where the suffixes $L$ and $R$ index two parts, $0\leq x\leq 0.2$ and $0.2<r\leq 1$, respectively. Here we choose $v_{a}=2.7$, $v_{b}=2.2$, $v_{c}=1.2$, $\eta_{a}=5$, $\eta_{b}=3$, $\eta_{c}=1.1$, $I=3$, $\Delta t=5\times10^{-6}$, $\Delta x=\Delta y=2\times10^{-4}$, $Q=1$. The collision parameters in MRT are $R_{5}=R_{6}=R_{7}=10^{4}$ and $10^{5}$ for the others.
\begin{figure}[tbp]
\begin{center}
\includegraphics[bbllx=7pt,bblly=316pt,bburx=560pt,bbury=608pt,width=0.9\textwidth]{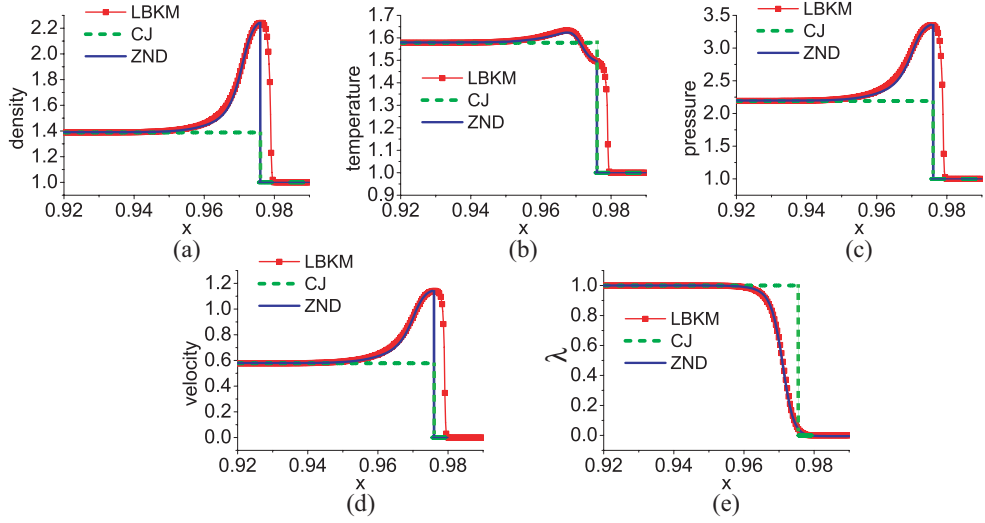}
\end{center}
\caption{(Color online) The profiles of the steady detonation: (a) $\rho $, (b) $T$, (c) $p$, (d) $u_{x}$, (e) $\lambda $.}
\label{Fig02}
\end{figure}
\begin{figure}[tbp]
\begin{center}
\includegraphics[bbllx=10pt,bblly=166pt,bburx=528pt,bbury=620pt,width=0.33\textwidth]{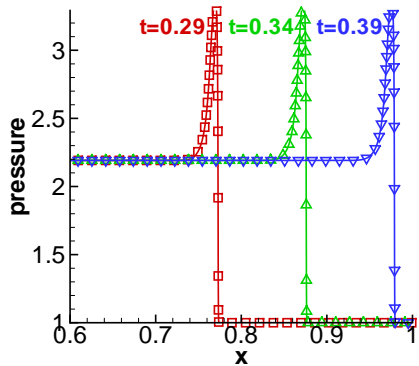}
\end{center}
\caption{(Color online) The profiles of $p$ in the evolution of the steady detonation at times $t=0.29$, $t=0.34$, $t=0.39$, respectively.}
\label{Fig03}
\end{figure}
Figure \ref{Fig02} shows the profile of the steady detonation at time $t=0.39$. Panels (a)-(e) gives physical quantities $\rho $, $T$, $p$, $u_{x}$, $\lambda $ versus $x$, respectively. The simulation results of LBKM, analytic solutions of Chapman-Jouguet (CJ) theory \cite{Chapman1899,Jouguet1905,Detonation2000} and
Zeldovich-Neumann-Doering (ZND) theory \cite{Zeldovich1940,Neumann1942,Doering1943,Detonation2000} are shown is each panel. The solid lines with squares, the dashed lines and the solid lines are for LBKM simulation results, CJ results and ZND results, respectively. The simulation results give physical quantities behind the detonation wave $(\rho ,T,u_{x},u_{\theta },\lambda)=(1.38869,1.57816,0.577384,0,1)$. Comparing them with CJ results gives the relative differences $0.023\%$, $0.025\%$, $0.006\%$, $0\%$ and $0\%$, respectively. It is clear in panels (a)-(e) that the LBKM simulation results agree well with the ZND results in the area behind von Neumann peak. But there exist significant difference in front of the von Neumann peaks. This is because the ZND theory used here ignores completely the effects of viscosity and heat conduction, and the von Neumann peak is treated simply as a strong discontinuity which is not true. While in the LBKM results the effects of viscosity, heat conduction and other kinds of relevant transportation are included. This difference will decrease with the decreasing of viscosity and heat conductivity. This point will be further discussed in section VI.

Figure \ref{Fig03} shows the pressure versus $x$ at times $t=0.29$, $t=0.34$, $t=0.39$, from left to right, respectively. Our simulation gives detonation velocity $v_D =2.06$, and the analytic solution is $v_D =2.06395$. The relative difference between them is $0.191\%$ which is satisfying.

\subsection{Riemann problems}

In this subsection our two-dimensional LBKM is used to solve the one-dimensional Riemann problems where there is no chemical reaction. Now we give simulation results for three typical Riemann problems, i.e., the Sod's shock tube, the Lax's shock tube and the Sjogreen's problem.

\subsubsection{Sod's shock tube}

\begin{figure}[tbp]
\begin{center}
\includegraphics[bbllx=4pt,bblly=425pt,bburx=581pt,bbury=542pt,width=1.0\textwidth]%
{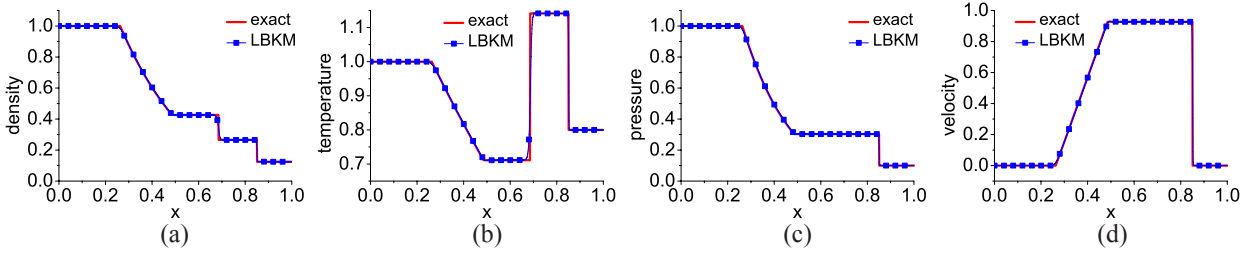}
\end{center}
\caption{(Color online) Comparison of numerical and theoretical results for the Sod shock tube at $t=0.2$. Solid lines are for exact solutions and solid lines with squares are for simulation results.}
\label{Fig04}
\end{figure}
For the problem of Sod's shock tube, the initial condition is described by
\begin{equation}
\left\{
\begin{array}{l}
(\rho ,T,u_{x},u_{y})_{L}=(1,1,0,0) \\
(\rho ,T,u_{x},u_{y})_{R}=(0.125,0.8,0,0)
\end{array}
\right.
\end{equation}
where left side $L\in [0,0.5)$ and the right side $R\in [0.5,1]$. Figure \ref{Fig04} shows the computed density, temperature, pressure, velocity profiles at the time $t=0.2$. The lines are for analytical solutions and solid lines with squares are for the LB simulation results. The size of grid is $\Delta x=\Delta y=10^{-3}$, time step $\Delta t=10^{-5}$, $I=3$ and ($v_{a}$, $v_{b}$, $v_{c}$, $\eta_{a}$, $\eta_{b}$, $\eta_{c}$) $=$ ($2.5$, $2.2$, $1.2$, $6.5$, $3$, $0$). The collision parameters in MRT are $R_{5}=R_{6}=R_{7}=1.2\times 10^{4}$, and other values of $R_{i}$ are $10^{5}$. It is easy to find in Fig.\ref{Fig04} that the two sets of results have a satisfying agreement.

\subsubsection{Lax's shock tube}

\begin{figure}[tbp]
\begin{center}
\includegraphics[bbllx=4pt,bblly=425pt,bburx=581pt,bbury=542pt,width=1.0\textwidth]{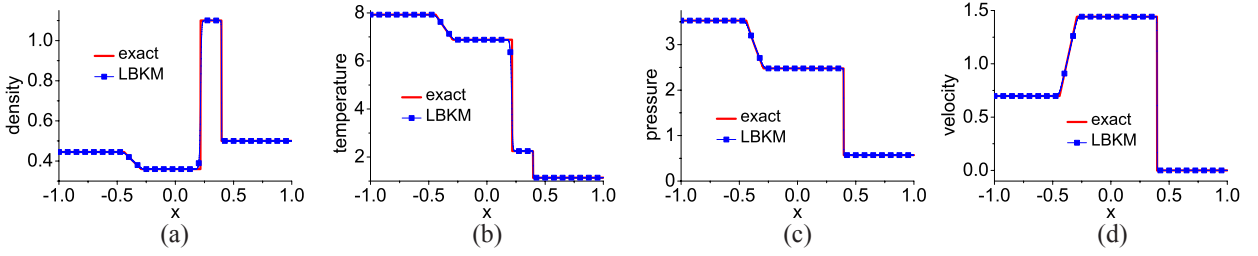}
\end{center}
\caption{(Color online)  Comparison of numerical and theoretical results for the Lax shock tube at $t=0.15$. Solid lines are for exact solutions and solid lines with squares are for simulation results.}
\label{Fig05}
\end{figure}
For this problem, the initial condition is described by
\begin{equation}
\left\{
\begin{array}{l}
(\rho ,T,u_{x},u_{y})_{L}=(0.445,7.928,0.698,0) \\
(\rho ,T,u_{x},u_{y})_{R}=(0.5,1.142,0,0)
\end{array}
\right.
\end{equation}
where $L\in [-1,0)$ and $R\in [0,1]$. Figure \ref{Fig05} shows the physical quantities (density, temperature, pressure, velocity) versus $x$ at the time $t=0.15$. The lines are for exact solutions and solid lines with squares correspond to our simulation results. The parameters are set to be $\Delta x=\Delta y=10^{-3}$, $\Delta t=10^{-5}$, $I=1$, ($v_{a}$, $v_{b}$, $v_{c}$, $\eta_{a}$, $\eta_{b}$, $\eta_{c}$) $=$ ($4.7$, $3.3$, $1$, $6$, $2.5$, $0.9$). The collision parameters in MRT are $R_{5}=R_{6}=R_{7}=2\times 10^{4}$, $R_{8}=R_{9}=8\times 10^{4}$, and other values of $R_{i}$ are $10^{5}$. We also find a good agreement between the exact solutions and our simulation results.

\subsubsection{Sjogreen's problem}

\begin{figure}[tbp]
\begin{center}
\includegraphics[bbllx=4pt,bblly=425pt,bburx=581pt,bbury=542pt,width=1.0\textwidth]{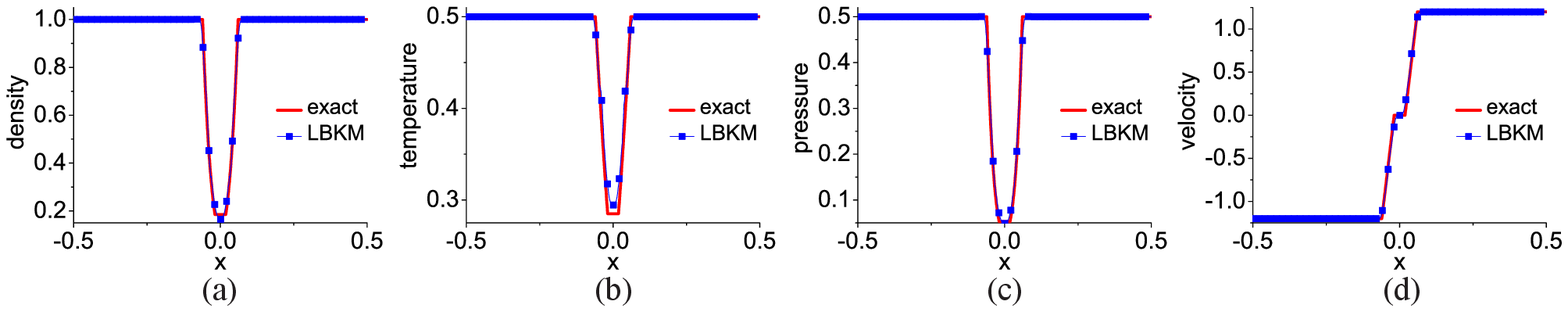}
\end{center}
\caption{(Color online)  Comparison of numerical and theoretical results for the Sjogreen's problem at $t=0.03$. Solid lines are for exact solutions and solid lines with squares are for simulation results.}
\label{Fig06}
\end{figure}
The initial condition for the Sjogreen's problem is
\begin{equation}
\left\{
\begin{array}{l}
(\rho ,T,u_{x},u_{y})_{L}=(1.0,0.5,-1.2,0) \\
(\rho ,T,u_{x},u_{y})_{R}=(1.0,0.5,1.2,0)
\end{array}
\right.
\end{equation}
where $L\in [-0.5,0)$ and $R\in [0,0.5]$. Figure \ref{Fig06} shows the physical quantities versus $x$ at the time $t=0.03$. The specific correspondences are referred to the legends. The parameters used here are $\Delta x=\Delta y=2\times 10^{-3}$, $\Delta t=2\times 10^{-5}$, $I=4$, ($v_{a}$, $v_{b}$, $v_{c}$, $\eta_{a}$, $\eta_{b}$, $\eta_{c}$) $=$ ($0.4$, $1.0$, $1.8$, $0.3$, $1.9$, $1.5$). The collision parameters in MRT are $R_{8}=R_{9}=2\times 10^{4}$, and others $5\times 10^{4}$. We also find a good agreement between the two sets of results.

\subsection{Shock reflection}

\begin{figure}[tbp]
\center
\includegraphics*[bbllx=22pt,bblly=209pt,bburx=510pt,bbury=580pt,angle=0,width=0.50\textwidth]{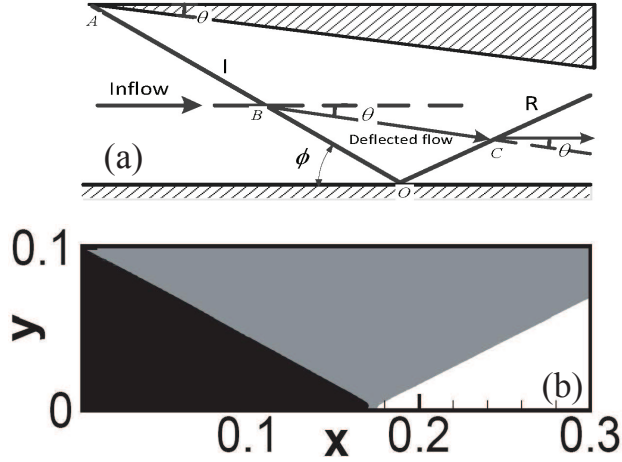}
\caption{Schematic (Fig. (a)) and density contour (Fig.(b)) of steady regular shock reflection on a wall. In (b), from black to white, the density increases.}
\label{Fig07}
\end{figure}
Shock reflection problem, which has been the subject of considerable
research effort over the last seven decades, is one of the most important
problems in both the science and engineering fields. Of particular interest
is, in general, the transition from so-called regular to irregular
reflection. Consider a plane shock (for example, one generated by a wedge in
steady invisid flow) being reflected off a wall, as schematically shown in
Fig.\ref{Fig07}(a). The type of reflection depends on ($M_{1}$, $\gamma $ , $\theta $)
parameter space, where $M_{1}$, $\gamma $ and $\theta $ are the incident
shock wave Mach number, gas specific heat ratio and flow deflection angle
respectively. In regular reflection the incident shock wave (I) and the
reflected shock wave (R) meet at the surface and is typical for a large wall
angle $\phi $. In the case of regular reflection, the conservations of mass,
momentum and energy relate the state downstream of the shock (subscript $2$)
to the state upstream (subscript $1$) as below:%
\begin{equation}
\frac{\rho _{2}}{\rho _{1}}=\frac{\left( \gamma +1\right) M_{1}^{2}\sin
^{2}\phi }{2+\left( \gamma -1\right) M_{1}^{2}\sin ^{2}\phi },  \label{RR1}
\end{equation}%
\begin{equation}
\frac{p_{2}}{p_{1}}=\frac{2\gamma M_{1}^{2}\sin ^{2}\phi -\left( \gamma
-1\right) }{\gamma +1},  \label{RR2}
\end{equation}%
\begin{equation}
M_{2}\sin ^{2}\left( \phi -\theta \right) =\frac{\gamma +1+\left( \gamma
-1\right) \left( M_{1}^{2}\sin ^{2}\phi -1\right) }{\gamma +1+2\gamma \left(
M_{1}^{2}\sin ^{2}\phi -1\right) },  \label{RR3}
\end{equation}%
\begin{equation}
\tan \theta =\frac{\tan \phi \left( M_{1}^{2}\cos ^{2}\phi -\cot ^{2}\phi
\right) }{1+\frac{1}{2}M_{1}^{2}\left( \gamma +\cos 2\phi \right) }.
\label{RR4}
\end{equation}%
For fixed $\gamma $ and $M_{1}$, the shock angle $\phi $ behaves as a
function of the deflection angle $\theta $.

Here we show a numerical test as below. An incoming shock wave with Mach
number $M_{1}=2.3094$ has an angle of $\phi =30^{\circ }$ to the wall. The
computational domain is a rectangle with length $0.3$ and height $0.1$. This
domain is divided into a $300\times 100$ rectangular grid with $\Delta
x=\Delta y=10^{-3}$. The boundary conditions are composed of a reflecting
surface along the bottom boundary, supersonic outflow along the right
boundary, and Dirichlet conditions on the left and top boundary conditions,
given by
\begin{equation}
\left\{
\begin{array}{lll}
(\rho ,T,u_{x},u_{y})_{0,y,t} & = & (1,0.5,2,0) \\
(\rho ,T,u_{x},u_{y})_{x,0.1,t} & = & (1.25,0.56,1.9,-0.173205)%
\end{array}%
\right.   \label{Boundary}
\end{equation}%
The parameters are chosen as $\Delta t=5\times 10^{-6}$, $I=2$ ($\gamma =1.5$%
), ($v_{a}$, $v_{b}$, $v_{c}$, $\eta _{a}$, $\eta _{b}$, $\eta _{c}$) $=$ ($%
1.0$, $2.7$, $2.9$, $1.0$, $2.9$, $0.96$). The collision parameters in MRT
are $R_{5}=R_{6}=R_{7}=1.8\times 10^{5}$, $R_{8}=R_{9}=2.0\times 10^{5}$,
and other values of $R_{i}$ are $10^{5}$. Figure \ref{Fig07}(b) shows contours
of density at $t=0.5$. The clear shock reflection on the wall agrees well
with the exact solution. (For example, from the boundary conditions, especially the bottom boundary condition shown by the second equation in Eq.(\ref{Boundary}), we obtain
 $\tan \theta =0.173205/1.9$. If substitute the values of $\phi $, $M_{1}$, $\gamma $ into Eq.(\ref{RR4}), we get exactly the same value for $\tan \theta$, $9.1161\times 10^{-2}$, if calculate in single-precision.)

\subsection{Couette Flow}
In order to demonstrate that the new model is also suitable for incompressible flows, we conduct a series of numerical simulations of Couette flow. The upper wall, with the distance $H=0.2$ apart from the lower wall, moves with a fixed speed $u_{0}$. The lower wall is at rest. Periodic boundary conditions are applied to the left and right boundaries, and the top and bottom adopt the non-equilibrium extrapolation method.

\begin{figure}[tbp]
\center\includegraphics*%
[bbllx=16pt,bblly=155pt,bburx=559pt,bbury=525pt,angle=0,width=0.30\textwidth]{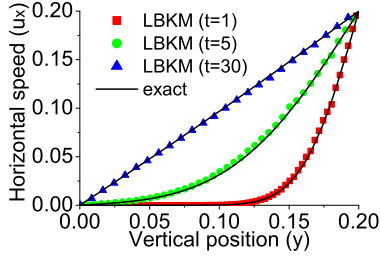}
\caption{(Color online) Horizontal speed distribution of Couette flow at various instants: $t=1$, $t=5$, $t=30$.}
\label{Fig08}
\end{figure}
In the first simulation of Couette flow, the initial state of the fluid is $\rho=1$, $T=1$, $u_{x}=u_{y}=0$. The viscous shear stress transmits momentum into the fluid and changes the horizontal speed profile \cite{Watari2003}. Figure \ref{Fig08} shows the horizontal speed distribution at various instants $t=1$, $5$, $30$. The simulation results coincide well with the following analysis,
\begin{equation}
u=\frac{y}{H}u_{0}+\frac{2}{\pi}u_{0}\sum_{n=1}^{\infty}%
[\frac{(-1)^{n}}{n}exp(-n^{2}\pi^{2}\frac{\mu t}{\rho H^{2}})sin(\frac{n\pi y}{H})]
\end{equation}
The parameters are $\Delta x=10^{-3}$, $\Delta t=10^{-5}$, $I=2$, ($v_{a}$, $v_{b}$, $v_{c}$, $\eta_{a}$, $\eta_{b}$, $\eta_{c}$) $=$ ($0.8$, $1.2$, $1.3$, $1.0$, $2.7$, $0.8$). The grid number is $N_{x}\times N_{y}=1\times 200$. The collision parameters are $R_{5}=R_{6}=R_{7}=2\times 10^{3}$, $R_{8}=R_{9}=1.0\times 10^{4}$, and the others $5\times10^{4}$.

\begin{figure}[tbp]
\center\includegraphics*%
[bbllx=4pt,bblly=326pt,bburx=581pt,bbury=541pt,angle=0,width=0.60\textwidth]{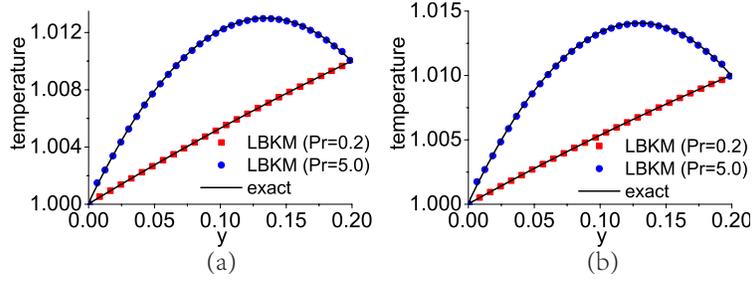}
\caption{(Color online) Temperature profiles of Couette flow. (a) $\gamma=1.4$. (b) $\gamma=1.5$}
\label{Fig09}
\end{figure}
Figure \ref{Fig09} shows the temperature profiles in another four simulations. In order to get a steady fluid state as soon as possible, we give the initial temperature field as below
\begin{equation}
T=T_{1}+(T_{2}-T_{1})\frac{x}{H}+\frac{\mu}{2\kappa} u_{0}^2 \frac{x}{H}(1-\frac{x}{H})
\end{equation}
where $T_{1}$ ($=1.0$) and $T_{2}$ ($=1.01$) are temperatures of the lower and upper walls, respectively. The initial velocity field is given as $u=u_{0} y/H$. And the time is $t=0.01$. Panels (a)-(b) correspond to $\gamma=1.4$ and $\gamma=1.5$, respectively. The case $Pr=0.2$ in panel (a) corresponds to $\Delta x=10^{-3}$, $\Delta t=10^{-5}$, $I=3$, $v_{a}=0.8$, $v_{b}=1.2$, $v_{c}=1.3$, $\eta_{a}=1.1$, $\eta_{b}=3.1$, $\eta_{c}=0.7$, $R_{5}=R_{6}=R_{7}=10^{4}$, $R_{8}=R_{9}=2\times10^{3}$, $R_{21}=R_{22}=R_{23}=R_{24}=10^{3}$, and $5\times10^{4}$ for other collision parameters. For the case $Pr=5.0$ in panel (a), the parameters are $\eta_{c}=0.7$, $R_{5}=R_{6}=R_{7}=2\times 10^{3}$, $R_{8}=R_{9}=10^{4}$, $R_{21}=R_{22}=R_{23}=R_{24}=5\times 10^{4}$, and the others are the same as those for $Pr=0.2$ in panel (a). Except $I=2$ and $\eta_{b}=2.1$, all the other parameters for the cases $Pr=0.2$ and $Pr=5.0$ in panel (b) are the same as those for the cases $Pr=0.2$ and $Pr=5.0$ in panel (a), respectively. It is clear that our simulation results are in agreement with the analytical ones, and the effects of the specific-heat ratio and the Prandtl number are successfully captured.

The analytical solutions used in the shock tube and shock reflection
problems are based on the Euler equations. The numerical tests show that, by
using large collision parameters (small viscosity and heat conductivity,
etc), the LBKM can present results having a satisfying agreement with those
based on the Euler equations.
The analytical solutions used in the Couette flow problems are based on the
Navier-Stokes equations. The numerical tests show that the LBKM can present
results having a satisfying agreement with those of Navier-Stokes equations.

\section{Physical gains and computing time}
We first discuss the computational costs for the MRT and SRT versions of
LBKM based on the same discrete velocity model. In 2013 we proposed a
uniform scheme for formulating LBKM\cite{XuGan2013}. In the current work we formulate the discrete velocity model according to the same idea. In this scheme, we first check which moment relations of $f^{eq}$
are needed to recover the hydrodynamic equations. Those moment relations can
be written in the uniform form,
\begin{equation}
\mathbf{Mf}^{eq}=\mathbf{\hat{f}}^{eq}\text{,}
\end{equation}%
where $\mathbf{M}$ is $N\times N$ matrix, $N$ is an integer to be fixed in
the next step. We rewritten the above moment relations in the explicit
Cartesian coordinates and check the equation number which is the value of $N$%
. In this way we can find the minimum number $N$ of needed discrete
velocities. This scheme works for both the MRT-LBKM and SRT-LBKM
formulations. In this formulation scheme, we can roughly estimate the
computational costs of the MRT and SRT versions as below.

The SRT and MRT versions of LB kinetic equations read
\begin{equation}
\frac{\partial f_{i}}{\partial t}+v_{i\alpha }\frac{\partial f_{i}}{\partial
r_{\alpha }}=-\frac{1}{\tau }\left( f_{i}-M_{il}^{-1}\hat{f}_{k}^{eq}\right) + C_{i}
\text{,}  \label{SRT}
\end{equation}%
and%
\begin{equation}
\frac{\partial f_{i}}{\partial t}+v_{i\alpha }\frac{\partial f_{i}}{\partial
r_{\alpha }}=-M_{il}^{-1}\left[ \hat{R}_{lk}\left( \hat{f}_{k}-\hat{f}%
_{k}^{eq}\right) +\hat{A_{l}}\right] + C_{i} \text{,}  \label{MRT}
\end{equation}%
respectively. Compared with the SRT version, the extra computation cost of
MRT includes two parts, the first part is for computing $\hat{f}%
_{k}=M_{ki}f_{i}$, the second is for $\hat{A_{l}}$ which has only two
non-zero terms. As mentioned in section II, an important skill here is that the inverse of the matrix $\mathbf{M} $ should be solved analytically before coding. We use the
software, MatLab2011, to do this. Thus, the elements of $\mathbf{M}^{-1}$
have been replaced by specific values before the main loop of the
simulation, instead of being numerically solved in each iteration
step.

To have a rough estimation on the computing time, we performed simulations of the same physical processes by using various LB models. The computational facility is a personal computer with Intel(R) Core(TM) $2$ CPU Q$9400$ @$2.66$GHz and RAM $4.00$ GB. Tabel I shows the computing times for three LBKM simulations of the same steady detonation behaviour. The first simulation in Table I is actually the first numerical test in Fig.
\ref{Fig02}. It is performed using the current MRT-LBKM with $24$ discrete velocities in Eq.(\ref{MRT}). The second is performed using the SRT-LBKM, (\ref{SRT}), with the same D2V24,
where the relaxation time is chosen as $\tau =10^{-5}$, and the other parameters are chosen as the same as the in first simulation. The third is performed using the
 SRT-LBKM described by the following equation,
\begin{equation}
\frac{\partial f_{i}}{\partial t}+v_{i\alpha }\frac{\partial f_{i}}{\partial
r_{\alpha }}=-\frac{1}{\tau }\left( f_{i}-f_{i}^{eq}\right)
 +C_{i}\text{,}
\label{BoltzmannSRT}
\end{equation}
  with the D2V33 by Watari\cite{Watari2003} where $33$ discrete velocities are used.
 Since the D2V33 works only for the case where the specific-heat ratio is fixed at $\gamma =2$, the parameter $I=0$ is used in third simulation, which does not influence the computing time.  It is easy to find that the computing time for the simulation using the current MRT is only $3\%$ more than that using the SRT with the same DVM, and is $7\%$ less than that using the SRT with DVM with $33$ discrete velocities.

\begin{center}
\begin{table}[tbp]
\label{TableCases}
\caption{Computing times for simulating a steady detonation process using various versions of the LBKM. }
\begin{tabular}{||c|c||}
\hline\hline
~~Model~~ & ~~Computing time~(unit: s)~~\\
\hline
MRT D2V24 & 2293.93 \\
\hline
SRT D2V24 & 2221.58 \\
\hline
SRT D2V33 & 2461.58 \\
\hline\hline
\end{tabular}
\label{ComputingTime}
\end{table}
\end{center}

It is interesting to have some comments on the MRT-LBKM versus the Navier-Stokes model.
\begin{enumerate}
\item The two-dimensional Navier-Stokes model is composed of $4$ nonlinear
partial differential equations. The current two-dimensional LBKM contains $24$
(formally) linear equations.

\item The linearity of the LB kinetic equations makes easy the algorithm and
coding. But the larger number of equations increase the computational cost.
If we are only interested in the density $\rho $, the momentum $\rho \mathbf{%
u}$ and the energy $E$, from which the flow velocity $\mathbf{u}$,
temperature $T$ can be obtained and then the pressure $p$ can also be obtained
from the equation of state, the Navier-Stokes model may be more efficient.

\item It is understandable that a lower-cost model is generally preferable.
A higher-cost model shows its necessity to be developed only in the
following two cases: (a) it can bring more information from which one can
gain a more complete or deeper insight into the problem under consideration,
or (b) it can bring more accurate results for the physical quantities under
consideration.

Physically, the proposed LBKM is roughly equivalent with a Navier-Stokes model
supplemented by a
coarse-grained model of the TNE behavious in the continuum limit. The
two-dimensional Navier-Stokes model describes the behaviours of the $4$
quantities, $\rho $, $\rho u_{x}$, $\rho u_{y}$ and $E$, which are conserved
in the collision process.
 The $20$ physical quantities, $\hat{f}_{k}-\hat{f}_{k}^{eq}$ (with $k=5$, $6$, $\cdots $, $24$%
),  constitute a rough description on the TNE behaviours.
The conserved and non-conserved
quantities are complimentary in describing more completely the behaviours
of complex flows. So, the LBKM proposed in the work belongs to the above
case (a).

As for case (b), by using the idea shown in this paper, it is
straightforward to construct a LBKM which can bring more accurate values of $%
\rho $, $\rho u_{x}$, $\rho u_{y}$ and $E$ than the Navier-Stokes model in the case or region where the local Knudsen number $\varepsilon $ is high, for example, around
strong shock/detonation waves or when the flow behaviour under consideration
is much faster than the case considered in this work. To that aim, we need
only use a new DVM constructed according to a longer list of moment
relations of $f^{eq}$. To save the computational cost, we can prepare
several, at least two, DVMs in the code. The DVM can be chosen adaptedly
according to the local Knudsen number $\varepsilon $. For example, when the
local Knudsen number $\varepsilon $ is higher than the case where the
Navier-Stokes model works, the code will adaptedly use a new DVM with more
discrete velocities. When the local Knudsen number $\varepsilon $ is smaller
than some critical value, the code will adaptedly use a DVM with fewer
discrete velocities. A coarse-grained modeling or approximation, $%
f_{i}=f_{i}^{eq}$, can be used at the first iteration step after switching
to a new DVM. The careful discussion on LBKM with flexible DVMs is out of
the scope of the paper.

\item One can always obtain the evolution equations of the non-conserved
quantities via the Chapman-Enskog analysis to the Boltzmann equation, which
is independent of the LBKM. That is to say, without LBKM, one can also solve
the coupled $24$ evolution equations of conserved and non-conserved physical
quantities using traditional CFD scheme. But solving the coupled $24$ nonlinear
partial differential equations is not an easy task. The $24$ LB kinetic
equations are (formally) linear and have the same form. The computations in LBKM are easy to be parallelized. In brief, when one aims to investigate both the HNE and TNE behaviours, the LBKM is a
convenient model.
\end{enumerate}

\section{non-equilibrium investigation of Detonation}

The LB kinetic model inherits naturally the function of Boltzmann equation, describing non-equilibrium effects in the system \cite{Succi-Book,ProgPhys2014,Review2012,XuYan2013,XuGan2013,XuLin2014PRE,XuLin2014CTP}. The departure of the system from local thermodynamic non-equilibrium can be measured by the difference between the high order kinetic moments of $f_{i}$ and $f^{eq}_{i}$ which are just ($\hat{f_{k}}-\hat{f}^{eq}_{k}$) in the current MRT-LB kinetic equation, (\ref{Boltzmann_chemical}).
We define
\begin{equation}
\Delta_{k}=\hat{f_{k}}-\hat{f}^{eq}_{k} \tt{.}
\end{equation}
It is easy to find that $\Delta_{k}=0$ for $k=1$, $2$, $3$, $4$ due to the conservation of mass, momentum and energy. Each non-zero $\Delta_{k}$ quantitatively describes the deviation status of the system from its local thermodynamic equilibrium from its own side. We can observe the thermodynamic non-equilibrium state in the $N$-dimensional space opened by $\Delta_{k}$ with $k=1$, $2$, $\cdots$, $N$. We further define a distance
 \begin{equation}
d = \sqrt{\sum^{N}_{1} \Delta_{k }^{2}} \tt{,}
\end{equation}
which is a rough and averaged estimation of the deviation amplitude from the thermodynamic equilibrium, where $\Delta_{k }$ is assumed to be dimensionless. Thus, $d=0$ when the system is in the thermodynamic equilibrium and $d >0$ in the thermodynamic non-equilibrium state.

In this part we give some results of $\Delta_{k}$ in the evolution of detonation. Corresponding to the simple definition of $\Delta_{k}$, we introduce some clear symbols as $\Delta_{v_{x}v_{x}}=\Delta_{5}$, $\Delta_{v_{x}v_{y}}=\Delta_{6}$, $\Delta_{v_{y}v_{y}}=\Delta_{7}$, $\Delta_{\eta^2}=\Delta_{4}-\Delta_{5}-\Delta_{7}$, $\Delta_{(v^2+\eta^2)v_{x}}=\Delta_{8}$, $\Delta_{(v^2+\eta^2)v_{y}}=\Delta_{9}$, $\Delta_{v_{x}v_{x}v_{x}}=\Delta_{10}$, $\Delta_{v_{x}v_{x}v_{y}}=\Delta_{11}$, $\Delta_{v_{x}v_{y}v_{y}}=\Delta_{12}$, $\Delta_{v_{y}v_{y}v_{y}}=\Delta_{13}$.

 A short discussion is as below. The non-equilibrium behaviours of various modes may contribute to the system evolution according to different amplification factors $R_{k}$, while all the amplification factors becomes the same in the SRT-LB model. Mathematically, the part $R_{k} \Delta _{k}$ in the right side of Eq.(\ref{CE_LB}) increases with increasing $R_{k}$ for fixed $\Delta_{k}$.

In this section we first investigate the unsteady detonation, then the compare cases where the detonation changes from unsteady to steady. All these cases show complex interplay between various HNE behaviours, between various TNE behaviours and between the HNE and TNE  behaviours.

\subsection{Unsteady detonation: simulations with different space and time steps}

Now we investigate some non-equilibrium behaviours in detonation phenomena. The initial physical quantities ($\rho ,T,u_{x},u_{y},\lambda$) are given the same values as those in Eq.(\ref{V_and_V}). Here we choose $v_{a}=2.7$, $v_{b}=2.2$, $v_{c}=1.2$, $\eta_{a}=1.5$, $\eta_{b}=0.5$, $\eta_{c}=5.0$, $I=3$, $Q=1$. The collision parameters in MRT are $R_{i}=100$. In numerical simulations the space and time steps should be small enough so that the spurious transportation behaviours are negligible compared with the physical ones. To assure that the numerical errors are small enough, we simulate the same detonation behaviour using three sets of spatial and temporal steps: (i) $\Delta x=\Delta y=10^{-3}$, $\Delta t=10^{-5}$; (ii) $\Delta x=\Delta y=10^{-3}$, $\Delta t=10^{-6}$; (iii) $\Delta x=\Delta y=10^{-4}$, $\Delta t=10^{-6}$.

\begin{figure}[tbp]
\center\includegraphics*%
[bbllx=7pt,bblly=52pt,bburx=575pt,bbury=821pt,width=0.9\textwidth]{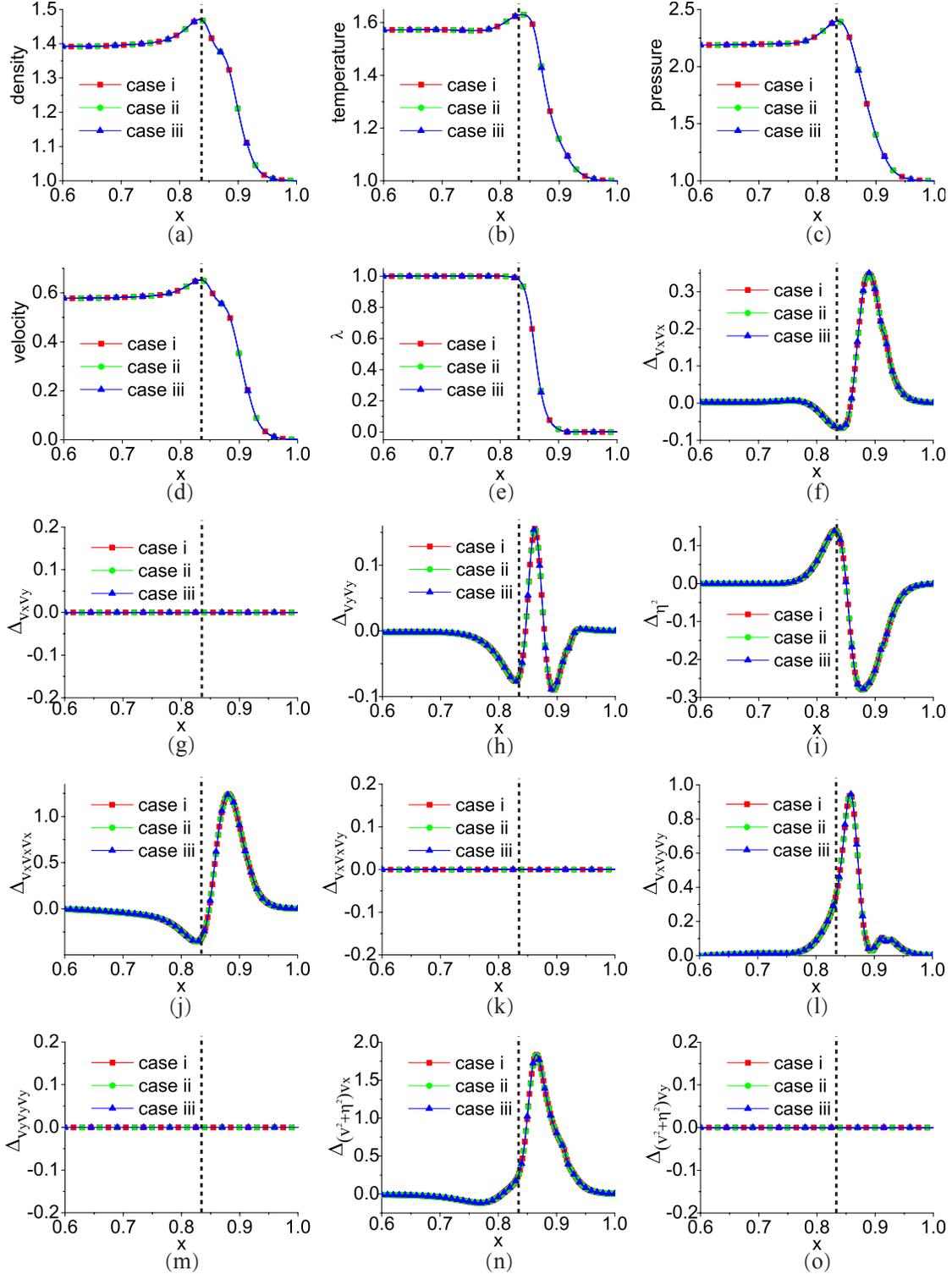}
\caption{(Color online) Physical quantities versus $x$ at time $t=0.35$ in three cases: (a) $\rho $, (b) $T$, (c) $p$, (d) $u_{x}$, (e) $\lambda$, (f) $\Delta_{v_{x}v_{x}}$, (g) $\Delta_{v_{x}v_{y}}$, (h) $\Delta_{v_{y}v_{y}}$, (i) $\Delta_{\eta^2}$, (j) $\Delta_{v_{x}v_{x}v_{x}}$, (k) $\Delta_{v_{x}v_{x}v_{y}}$, (l) $\Delta_{v_{x}v_{y}v_{y}}$, (m) $\Delta_{v_{y}v_{y}v_{y}}$, (n) $\Delta_{(v^2+\eta^2)v_{x}}$, (o) $\Delta_{(v^2+\eta^2)v_{y}}$.}
\label{Fig10}
\end{figure}
Figure \ref{Fig10} shows the simulation results of physical quantities ($\rho $, $T$, $p$, $u_{x}$, $\lambda$, $\Delta_{v_{x}v_{x}}$, $\Delta_{v_{x}v_{y}}$, $\Delta_{v_{y}v_{y}}$, $\Delta_{\eta^2}$, $\Delta_{v_{x}v_{x}v_{x}}$, $\Delta_{v_{x}v_{x}v_{y}}$, $\Delta_{v_{x}v_{y}v_{y}}$, $\Delta_{v_{y}v_{y}v_{y}}$, $\Delta_{(v^2+\eta^2)v_{x}}$, $\Delta_{(v^2+\eta^2)v_{y}}$) versus $x$ at time $t=0.35$ in the three cases. A vertical dashed guideline is plotted in each panel to guide the eye for the horizontal position $x=0.8345$ corresponding to the peak of pressure. It should be pointed out that up to this time the detonation has not obtained its steady state. The pressure at the von-Neumann-peak will increase further. We choose such a time because it is interesting to study the complex interplay between various non-equilibrium behaviours in the unsteady detonation process.

(1) At the same time, $t=0.35$, the detonation shown in Fig. \ref{Fig03} has already been steady, but the current one has not. The physical reason is that the viscosity of the system here is much larger than that shown in Fig.\ref{Fig03}. It takes more time for the steady detonation wave to form.

(2) All the simulation results in panels (a)-(o) are physically reasonable.
The simulation results of each quantity in the three cases have a satisfying
coincidence. It shows that the grid size $0.001$ and the time step $10^{-5}$ are
small enough for the current problem.
Given $R_{i}$ small enough, the physical viscosity is much larger than numerical viscosity here.

(3) Panels (a)-(e) show that the maximum values of density, temperature, pressure, velocity are not located at the same $x$-coordinate, and the pressure peak is located in the reaction zone where $0<\lambda<1$. It means that, before the reaction finishes, the temperature first arrives at its peak value, then the pressure, density and flow velocity arrives at their peak values. Here we refer the von-Neumann-peak as to the point where the pressure has its largest value. When the reaction finishes, all the density, temperature, pressure and the flow velocity have passed their peak values.

(4) As shown in panels (f), (h) and (i), the simulation results of $\Delta_{v_{x}v_{x}}$, $\Delta_{v_{y}v_{y}}$ and $\Delta_{\eta^2}$ satisfy the relation $\Delta_{v_{x}v_{x}}+\Delta_{v_{y}v_{y}}+\Delta_{\eta^2}=0$. Here what $\Delta_{v_{x}v_{x}}$, $\Delta_{v_{y}v_{y}}$ and $\Delta_{\eta^2}$ describe are the departures of the internal energies in the $x$, $y$ and internal degrees of freedom from their average. The relaxation coefficients $R_{5}$ and $R_{7}$ are related to evolution speeds of the internal energies in $x$ and $y$ degree of freedom, respectively. This result is physically reasonable. The results in  panels (f), (h) and (i) show clearly that,
when the system is not in its thermodynamic equilibrium state, the internal energies in different degrees of freedom may not equal each other, that the exchange of the internal energies in different degrees of freedom, due to the molecule collision, makes them evolve  towards their average.

(5) Both $\Delta_{v_{x}v_{x}}$ and $\Delta_{\eta^2}$ show a crest and a trough in the reaction zone, while $\Delta_{v_{y}v_{y}}$ shows a crest and two troughs. The result of $\Delta_{v_{x}v_{x}}$ first shows a crest and then a trough when the detonation wave travels forward. While $\Delta_{\eta^2}$ show an opposite behaviour. The crest of $\Delta_{v_{y}v_{y}}$ is in between its two troughs. Physically, comparing with the internal energy in the $y$ or extra degree of freedom, the internal energy in the $x$ degree of freedom first increases in the preshocked area. The maximum absolute value of $\Delta_{v_{x}v_{x}}$ is the largest among $\Delta_{v_{x}v_{x}}$, $\Delta_{v_{y}v_{y}}$ and $\Delta_{\eta^2}$ in the whole range shown in the figure.

(6) Panels (g), (k), (m), (o) show that the results of $\Delta_{v_{x}v_{y}}$, $\Delta_{v_{x}v_{x}v_{y}}$, $\Delta_{v_{y}v_{y}v_{y}}$, $\Delta_{(v^2+\eta^2)v_{y}}$ are equal to zero. Here $\Delta_{v_{x}v_{y}}$ associates with the shear effect, $\Delta_{v_{x}v_{x}v_{y}}$, $\Delta_{v_{y}v_{y}v_{y}}$, $\Delta_{(v^2+\eta^2)v_{y}}$ are related to ``the internal energy flow caused by microscopic fluctuation" in $y$ direction. The results are consistent with the fact that the simulated system is one dimensional or uniformly symmetric in the $y$ direction. There is neither shear effect nor energy flux in the $y$ direction.

(7) It can be found in panels (j), (l) and (n) that $\Delta_{v_{x}v_{x}v_{x}}$, $\Delta_{v_{x}v_{y}v_{y}}$, $\Delta_{(v^2+\eta^2)v_{x}}$ deviate significantly from zero. $\Delta_{v_{x}v_{x}v_{x}}$, $\Delta_{v_{x}v_{y}v_{y}}$, $\Delta_{(v^2+\eta^2)v_{x}}$ are associated with ``the internal energy flow caused by microscopic fluctuation" in the $x$ direction. As the chemical energy is released continuously in the reaction zone, the compression and rarefaction make effects on the system successively, Those actions make the velocity distribution function asymmetrical about the point ($v_{x}=u_{x}$, $v_{y}=u_{y}$) which is the symcenter of the Maxwellian distribution (see Eq. (\ref{feq_continuous})). Consequently, $\Delta_{v_{x}v_{x}v_{x}}$, $\Delta_{v_{x}v_{y}v_{y}}$ deviate from zero in the reaction zone.

\subsection{Detonations from unsteady to steady: simulations with different collision parameters}

Now we study detonation phenomena with different collision parameter sets: (i) $R_{i}=10^{2}$, (ii) $R_{i}=10^{3}$, (iii) $R_{i}=10^{4}$. The first case here is just the case (i) in the above subsection. In the second case where $R_{i}=10^{3}$, the parameters
($v_{a}$, $v_{b}$, $v_{c}$, $\eta_{a}$, $\eta_{b}$, $\eta_{c}$)
=($2.7$, $2.6$, $1.9$, $5.0$, $0.0$, $1.7$), $\Delta x=2\times10^{-4}$, $\Delta t=2\times10^{-6}$, the other parameters are the same as the first case.
In the third case where $R_{i}=10^{4}$, the spacial and temporal steps $\Delta x=4\times10^{-5}$, $\Delta t=4\times10^{-6}$, the other parameters are the same as the second case.

\begin{figure}[tbp]
\center\includegraphics*%
[bbllx=7pt,bblly=0pt,bburx=575pt,bbury=302pt,width=0.9\textwidth]{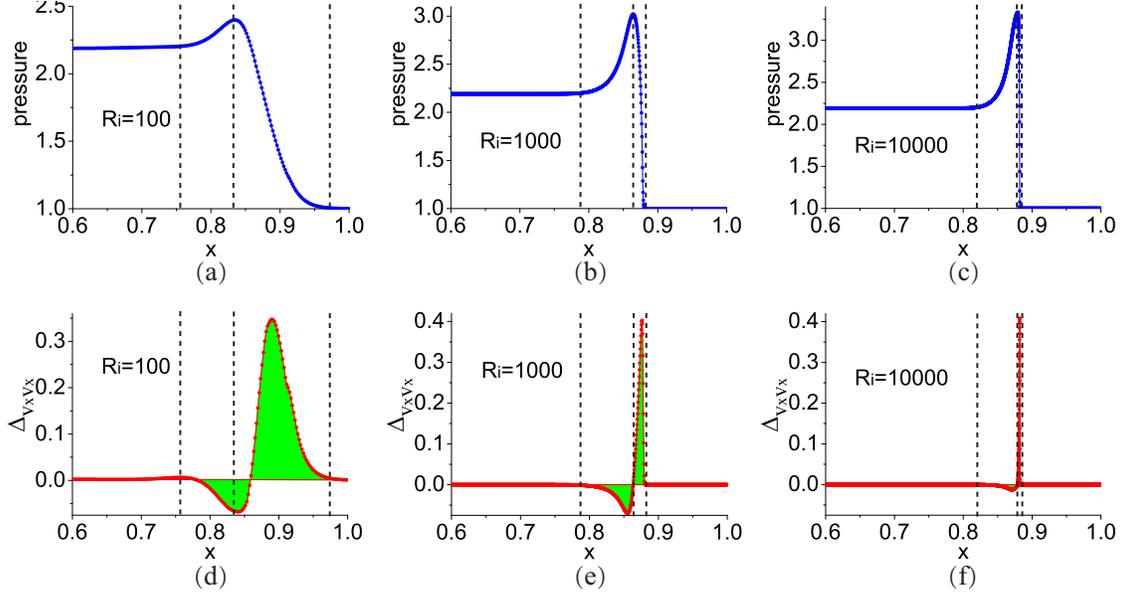}
\caption{(Color online) The physical quantities $p$ and $\Delta_{v_{x}v_{x}}$ versus $x$ at time $t=0.35$. The first row is for the pressure and the second is for $\Delta_{v_{x}v_{x}}$. From left to right, the three columns are for $R_{i}=10^{2}$, $R_{i}=10^{3}$ and $R_{i}=10^{4}$, respectively.}
\label{Fig11}
\end{figure}
Figure \ref{Fig11} shows the simulation results of $p$ and $\Delta_{v_{x}v_{x}}$ versus $x$ at time $t=0.35$ in the three cases with $R_{i}=10^{2}$, $R_{i}=10^{3}$ and $R_{i}=10^{4}$, respectively. Here we define ($X_{m}$, $P_{m}$) as the point where the largest pressure is located. The points in panels (a)-(c) are ($0.83450$, $2.39850$), ($0.86410$, $3.01965$) and ($0.87906$, $3.33212$), respectively. A vertical dashed guideline is plotted across this point in each panel. Aside from this guideline, another two lines are given to guide the eye for the width of the detonation wave. At the right side of the rightmost line is the unreacted explosive in metastable equilibrium with zero reaction rate. At the left side of the leftmost line, where all the materials are reaction products, the system is in a constant state. From (a) to (c), the detonation wave at this time changes from unsteady to steady. It is interesting to study the TNE behaviours in these cases. The guidelines in panels (d)-(f) coincide with those in (a)-(c), respectively.

(1) It is clear to find in panels (a)-(f) that the detonation wave, especially the preshocked area, becomes narrower with increasing $R_{i}$. Physically, the viscosity which is inversely proportional to $R_{\mu}$ widens the detonation wave, especially the preshocked area. Correspondingly, the area of nonequalibrium system is widened as well.

(2) Panels (a)-(c) show that both $X_{m}$ and $P_{m}$ increase from left to right. That is to say, with the increase of $R_{i}$, it takes less time for the detonation to become steady, and the von Neumann peaks becomes higher and sharper. Physically, the viscosity expands and smoothes the wave front of pressure. Consequently, it decreases the local TNE effects.

(3) In panels (d)-(f), the shaded area enclosed by the curve $\Delta_{v_{x}v_{x}}(x)$ and the line $\Delta_{v_{x}v_{x}}=0$ decreases from left to right. This shaded area presents a rough description on the global TNE effect around the detonation wave in the system. From this aspect, the viscosity increases the global TNE effect.

(4) The minimum of $\Delta_{v_{x}v_{x}}$ is $-0.06759$, $-0.07018$, $-0.01275$ in panels (d)-(f), respectively. The corresponding maximum is $0.34753$, $0.40275$, $0.40857$, respectively. The minimum of $\Delta_{v_{x}v_{x}}$ for $R=1000$ is less than the other two, and the maximum for $R=10000$ is the largest among the three cases. There is competition between the viscosity/(heat conductivity) effect and the gradient effects of physical quantities ($\rho$, $\mathbf{u}$, $T$, $p$, etc.). With the increase of collision parameters, the viscosity and heat conductivity decrease, while the gradients of physical quantities increase. The former tend to decrease and the latter tend to increase the TNE effects. The physical reason is that the viscosity possesses both the thermodynamic and hydrodynamic functions. Thermodynamically, it tends to make the system approach the thermodynamic equilibrium more slowly. But hydrodynamically, it works as a kind of resistance force to the shocking process, makes the pressure curve smoother, and consequently tends to make the system approach the thermodynamic equilibrium more quickly.
The heat conductivity plays a similar role.

(5) The first horizontal position for $\Delta_{v_{x}v_{x}}=0$ behind the von Neumann peak
    moves towards the horizontal position for the von Neumann peak as $R_{i}$ increases. It can be found a clear distance from the position for  $\Delta_{v_{x}v_{x}}=0$ to the position for the von Neumann peak in panel (a). While they almost coincide in (c). That is to say, the position, where the internal energy in the $x$ degree of freedom equals to the average of all degrees of freedom, gets away from the position for the von Neumann peak with increasing viscosity.

\section{Conclusion and discussions}

A MRT-LBKM for combustion phenomena is presented. The chemical energy released in the progress of combustion is dynamically coupled into the physical system by adding a chemical term to the LB kinetic equation. The chemical term describes the change rate of distribution function $f$ due to the local chemical reaction. Physically, the new model is roughly equivalent with a Navier-Stokes model supplemented by a coarse-grained model of the thermodynamic non-equilibrium behaviours in the continuum limit.
In this model the discrete equilibrium distribution function $f^{eq}_{i}$ needs to satisfy $24$ independent kinetic moment relations. We present a new discrete velocity model with $24$ velocities which are divided into $3$ groups. In each group a flexible parameter ($v_{a}$, $v_{b}$ or $v_{c}$) is used to control the size of discrete velocities and a second parameter ($\eta_{a}$, $\eta_{b}$ or $\eta_{c}$) is used to describe the contribution of the extra degrees of freedom. The current model works for both subsonic and supersonic flows with or without chemical reaction.
The rate equation for the chemical reaction can be adjusted according to specific situations.
In the MRT-LBKM, the non-equilibrium behaviours of various modes may contribute to the system evolution according to different amplification factors.

As an initial application, various non-equilibrium behaviours around the detonation wave in one-dimensional detonation process are preliminarily probed.
The following thermodynamic non-equilibrium behaviours, (i) exchange of internal kinetic energy between different degrees of freedom for molecule displacements, (ii) exchange of internal kinetic energy between the displacements and the internal degrees of freedom of the molecules,
are observed.
  It is found that the system viscosity (or heat conductivity)  decreases the local thermodynamic non-equilibrium, but increase the global thermodynamic non-equilibrium around the detonation wave, that even locally, the system viscosity (or heat conductivity)  results in two kinds of competing trends, to increase and to decrease the thermodynamic non-equilibrium effects.
The physical reason is that the viscosity (or heat conductivity)  takes part in both the thermodynamic and hydrodynamic responses to corresponding driving forces.
When we consider the thermodynamic non-equilibrium which can be described by various kinetic moments of $f-f^{eq}$, the Boltzmann equation (\ref{CE_LB})
can be regarded as a kind of constitutive equation relating to the response,
$\hat{\mathbf{f}}-\hat{\mathbf{f}}^{eq}$, to the driving force, $-\partial
\hat{\mathbf{f}}/\partial t-\partial (\hat{\mathbf{E}}_{\alpha }\hat{\mathbf{%
f}})/\partial r_{\alpha }+\hat{\mathbf{A}}+\hat{\mathbf{C}}$. Thus, the
inverse of the collision parameter, $\hat{\mathbf{R}}^{-1}$, plays a role of
the parameter describing material kinematic property. Thermodynamically, it
tends to amplify the thermodynamic non-equilibrium effects. But hydrodynamically, the viscous force tends to decrease the pressure gradient, the heat conduction tends to decrease the temperature gradient, and consequently they tend to decrease thermodynamic non-equilibrium.

If the local temperature increment due to chemical reaction is dynamically taken into account in the calculation of local equilibrium distribution function $f^{eq}_{i}$\cite{XuYan2013}, the number of needed discrete velocities can be decreased to $16$ in the two-dimensional case. In that case, only $12$ non-conserved quantities are included in the two-dimensional MRT-LBKM.

\section*{Acknowledgements}

The authors thank the anonymous referees and Prof. Sauro Succi for helpful comments and suggestions on improving the manuscript, thank Profs. Zheng Chen, Yanbiao Gan and Huilin Lai for fruitful discussions on modeling combustion and complex fluids.
AX and GZ acknowledge support of the Science Foundations of National Key Laboratory of
Computational Physics, National Natural Science Foundation of China [under Grant Nos. 11475028 and 11202003],  the opening project of State Key Laboratory of Explosion Science and Technology (Beijing Institute of Technology) [under Grant No. KFJJ14-1M] and the Open Project Program of State Key Laboratory of Theoretical Physics, Institute of Theoretical Physics, Chinese Academy of Sciences, China [under Grant No. Y4KF151CJ1]. YL and CL acknowledge support of National Natural Science Foundation of China [under Grant Nos. 11074300 and 41472130], National Basic Research Program of China [under Grant No. 2013CBA01504].


\end{document}